\documentstyle[amssymb,aps,prb,manuscript]{revtex}
\tighten

\begin{document}
\draft
\title{Wave functions in disordered wires in a weak magnetic field}
\author{A.~V.~Kolesnikov$^{1}$and K.~B.~Efetov$^{1,2}$}
\address{$^{1}$Fakult\"{a}t f\"{u}r Physik und Astronomie, Ruhr-Universit\"{a}t\\
Bochum, Universit\"{a}tsstr. 150, Bochum, Germany\\
$^{2}$L.D. Landau Institute for Theoretical Physics, Moscow, Russia}
\date{\today{}}
\maketitle

\begin{abstract}
Using the supersymmetry technique combined with the transfer matrix approach
we calculate different physical quantities characterizing localization in
disordered wires. In particular, we analyze the density-density correlation
function and its moments and study effects of an external magnetic field $H$
on tails of wave functions. At zero and very strong magnetic fields, we
obtain explicit expressions for all moments and for the entire distribution
of the wave functions. The crossover between the two limiting cases is more
difficult and calculations are performed for weak magnetic fields only. We
found that the far tail of the average density-density correlation function
and of its moments is strongly influenced by a weak magnetic field and
decays twice as slow as their main body. Extending Mott's physical picture
for the localized states we present also a qualitative description of the
crossover in the magnetic field. From both the analytical calculations and
the qualitative description we argue that the slower decay of the averaged
quantities is a consequence of rare but large splashes of wave functions at
large distances. The distribution of the logarithm of the wave function 
should also be affected by the magnetic field. The splashes and the 
two-scale decay of the averaged correlation functions can be relevant for 
the conductance of a system of disordered wires connected in parallel.
\end{abstract}


%\large

%\begin{multicols}{2} 

\section{Introduction}

Localization of all states in one-dimensional chains and
quasi-one-dimensional wires with a finite thickness is a very well known
phenomenon in the theory of disordered metals. Following the first works 
\cite{mott,thouless} where the localization was predicted, a number of
analytical methods has been developed to treat both the chains \cite
{Ber,abrikosov,gorkov} and wires \cite{EL,efetov,D} quantitatively. A
detailed discussion of the localization in the chains and wires can be
found, e.g. in Refs. \cite{gogolin,beenak,kram,book}

The localization of electron waves in weakly disordered chains and wires
occurs due to destructive interference and can be destroyed at finite
temperatures by inelastic scattering. The phenomenon of the electron
localization in thick wires is richer than in chains. The localization
length $L_{c}$ in chains is of the order of the mean free path $l$.
Electrons move at distances smaller than the localization length
ballistically without being scattered by impurities and get localized at
distances exceeding $L_{c}$. In contrast, the localization length $L_{c}$ in
wires is larger than the mean free path $l$ by a factor proportional to the
number of channels of the transversal quantization $N$. At distances larger
than $l$ but smaller than $L_{c}$, the electrons diffuse and become
localized only at distances larger than $L_{c}$.

One more important difference between chains and wires manifests itself if
an external magnetic field $H$ is applied. In chains the magnetic field
cannot influence the electron motion, whereas the motion in thick wires is
quite sensitive to it. It turns out that the magnetic field does not destroy
the localization of the wave functions but changes the localization length.
Remarkably, the localization length $L_{c}$ in thick wires exactly doubles
when applying a sufficiently strong magnetic field. \cite{EL,efetov}

For analytic calculations for thick wires a supermatrix $\sigma $-model \cite
{efetov,book} proved to be a powerful tool that allows one to reduce the
calculation of kinetic quantities at arbitrary frequencies $\omega $ to
solving of a system of differential equations. The localization length $%
L_{c} $ is obtained from an exponential decay of the density-density
correlation function. The most interesting limit $\omega \rightarrow 0$ can
be investigated explicitly and one obtains relatively simple formulae for
different physical quantities proving the localization of all states.

Calculation of the transmittance of a finite sample with ideal leads is the
basis of many numerical investigations of the localization properties. The
equivalence of different methods of calculation of the localization length $%
L_{c}$ is traditionally guaranteed by the Borland conjecture \cite{Borland}
stating a complete independence of the localization properties of the
boundary conditions.

The limiting cases of a strong magnetic field and no magnetic field, the
unitary and the orthogonal ensembles, respectively, are relatively well
studied in quasi 1D, whereas the crossover between these two limits has not
been well understood. It was assumed that the localization length changed
smoothly \cite{bouchaud,LerIm} between $L_{c}$ and $2L_{c}$ and one had to
calculate a curve connecting $L_{c}$ and $2L_{c}$ when increasing the
magnetic field from $0$ to $\infty $.

At the same time, investigation of the crossover can have quite interesting
applications now because a systematic experimental study of the localization
has been performed recently. In the experiment, \cite{Exp1,Exp2} a large
number of submicrometer-wide wires prepared from Si $\delta $-doped GaAs
were connected in parallel to insure statistical averaging. This enabled
accurate measurements of the conductivity as a function of the temperature $%
T $ and the magnetic field $H$ applied perpendicular to the wires. The
regimes of the weak and strong localization were observed and the
experimental data in the regime of strong localization were used to extract
the dependence of the localization length $L_{c}$ on $T$ and $H$. The
localization length $L_{c}$ was shown to double in a strong magnetic field
as compared with the limit of zero magnetic field $H$, which is in a
complete agreement with the theory. \cite{EL,D} As the measurements are done
at an arbitrary magnetic field, study of the crossover between the
orthogonal and unitary ensembles may become very important.

In a recent publication, \cite{WE} we discovered that already an arbitrarily
weak magnetic field drastically changes the tails of the averaged amplitudes
of the wave functions. At sufficiently large distances depending on the
magnetic field, the exponential decay with the localization length $L_{c}$
changes to a decay with the length $2L_{c}$. This result is of interest not
only from the theoretical point of view but it might also be important for
the transport measurements, \cite{Exp1,Exp2} since the localization length $%
L_{c}$ is relevant for the hopping conductivity at low temperature. \cite
{kurk,larkin}

However, a numerical study of the averaged logarithm of the transmittance
performed for sufficiently long wires \cite{Been-Sch} did not manifest any
two-scale behavior showed a smooth variation of the localization length
between $L_{c}$ and $2L_{c}$. No sharp change in the far tail of the
logarithm of the transmittance was observed in a weak magnetic field. As a
possible explanation of the contradiction several reasons were suggested. As
one of the possibilities, the difference between the logarithm of the
averaged wave functions, the quantity we have considered, \cite{WE} and the
averaged logarithm of the wave functions was mentioned. \cite{Been-Sch}

In this paper, we give details of the calculations presented in our short
publication \cite{WE} and discuss the question in what situations our effect
of the changing of the tails of the wave functions by a weak magnetic field
can be identified. To understand better the shape of the wave functions we
calculate not only the density-density correlation function but also its
moments. The changes in the tails of the wave functions are seen in all
moments of the density-density correlation functions and all of them are
proportional at large distances $x$ to $\exp (-|x|/4L_{{\rm cu}})$, where $%
L_{{\rm cu}}$ is the localization length for the unitary ensemble. The
independence of the exponential decay of the asymptotics of the number of
the moment signals that they are formed by strong rare splashes of the wave
functions. 

The situation with the averaged logarithm of the wave function is more
delicate. The large splashes contribute also to the averaged logarithm of the
wave functions at finite distances. As a result, some broad distribution of
logarithms occurs. Unfortunately, we are not able to predict how this 
distribution function develops in the limit of infinitely long distances. 
However, our results provide us reasons enough to suggest that the 
logarithmically normal distribution implied within the Borland conjecture 
is not universally applicable for length of the sample $x \gg L_c$ for the 
description of the crossover region. Existing numerical studies  
\cite{Pich1,Pich2,Been-Sch} might be interpreted as the head of some 
logarithmically normal distribution changes smoothly with $H$. Our results 
do not exclude a smooth behavior of the average logarithm of the wave 
function. We show, however, that another length scale, $x_H\sim L_c 
\ln(1/H)$ enters the theory of localization, up to which the 
distribution should differ from the standard logarithmically normal one.
Thus, the whole distribution function of the logarithm is of interest.

For calculations we use the supersymmetry technique that makes possible
consideration of disordered wires with any kind of boundary conditions. In
particular, one can study the transmittance of the disordered wires with
metallic leads. \cite{Zirn,Ziall} An external magnetic field is naturally
accounted for within supersymmetry formalism allowing study of the crossover
between the unitary and orthogonal ensembles.

The localization of the wave functions discussed in the present paper is
extracted from the moments of the density-density correlation function 
\begin{equation}
p_{\infty }^{(n)}\left( x-x^{\prime }\right) =\left\langle \sum_{\alpha
}\left| \psi _{\alpha }\left( x\right) \right| ^{2n}\left| \psi _{\alpha
}\left( x^{\prime }\right) \right| ^{2n}\delta \left( \varepsilon
-\varepsilon _{\alpha }\right) \right\rangle \;  \label{00}
\end{equation}
where $\psi _{\alpha }\left( x\right) $ and $\varepsilon _{\alpha }$ are the
eigenfunction and eigenenergy of a state $\alpha $, $x$ and $x^{\prime }$
are the coordinates along the wire, and the angle brackets stand for the
averaging over the disorder. This quantity was previously studied for a
closed finite multichannel wire, \cite{Fyo-Mir} when the wave functions were
taken at the ends of the wire. Generalizing the expressions for arbitrary $n$%
, the complete distribution was also obtained. \cite{Fyo-Mir}

Naturally, the very presence of hard boundaries or ideal leads inevitably
changes localized states near the ends of the disordered sample. These
states should differ from those localized in the bulk. Such a difference has
been found indeed for the density of states of a single chain. \cite{AltPrig}

The localization properties extracted from the transmittance of an open
system \cite{1chain1,1chain2} and from the density-density correlator of a
closed one, \cite{Fyo-Mir} manifest a generic universality: All moments of
the both quantities decay with the same rate, $\exp (-|x-x^{\prime
}|/4L_{c}) $, and the logarithm of the both quantities is normally
distributed. The fact that the decay rate is the same for all moments shows
that an important contribution comes from rare strong splashes of the wave
functions. Due to these splashes the moments decay slower with the length $%
4L_{c}$ and not with the localization length $L_{c}$. At the same time, the
logarithm of physical quantities like the transmittance or the product of
the amplitudes of the wave functions at different points is a self-averaging
quantity and approaches the value $-|x-x^{\prime }|/L_{c}$ for $|x-x^{\prime
}|\rightarrow \infty $. This quantity characterizes typical wave functions.
The question what happens with the wave functions if a weak magnetic field
is switched on has not been discussed.

In the subsequent chapters we present details of our analysis. We use an
approach based on the Lebedev-Kontorovich transformation to find the exact
distribution of any quantity of interest in quasi-1D wires for the
orthogonal and unitary ensemble. This allows us to calculate the correlator,
Eq.~(\ref{00}), and the Landauer-like conductivity. Exact expressions valid
at arbitrary distances are obtained for all moments for the pure orthogonal
and unitary ensemble. Using the formulae for the moments we derive the
entire distribution function. At a weak magnetic field, we show that a
peculiar two-scale behavior characterizes any moment of the wave functions,
Eq.~(\ref{00}).

The paper is organized as follows: In Sec. II we introduce the quantities to
be studied. A transfer-matrix formalism used for the calculations is
described in Sec. III. Reduction to an effective Schr\"{o}dinger equation is
carried out in Sec. IV. Then, in Sec. V we calculate the moments and the
distribution functions in the limiting cases of the orthogonal and unitary
ensembles. In Sec. VI we present results for the wire in a weak magnetic
field. Using a standard qualitative picture for localized states \cite
{Mott2,Mott3} we discuss the physics involved in Sec. VII. Our arguments
allow us to obtain without any calculation the characteristic length scales
of the problem and reach a better understanding of the localization in the
magnetic field. We discuss the results and make conclusions in Sec. VIII.

\section{Correlation Functions}

Localization in disordered metals can be well characterized by different
correlation and distribution functions. For an infinite sample one can
consider the density-density correlation function and its moments. For a
finite wire connected to metallic leads the transmittance and its moments
are of interest.

These quantities can be efficiently studied using the supersymmetry method. 
\cite{efetov,book} Following this approach one should reduce calculation of
the correlation functions of interest to computation of correlation
functions of a supermatrix $\sigma $-model. In the standard formulation the $%
\sigma $-model contains $8\times 8$ supermatrices $Q$. Fluctuations of the
supermatrices $Q$ are strongly influenced by a magnetic field and one half
of the ``degrees of freedom'' is suppressed if the magnetic field is
sufficiently strong. This corresponds to reducing the size of the
supermatrices $Q$ to $4\times 4$. Keeping this size of the supermatrices is
sufficient to calculate such physical quantities as conductivity or
density-density correlation function.

In order to calculate higher moments of these quantities or distribution
functions one has to increase the size of the supermatrices, which would
make the theory very complicated. Fortunately, calculation of moments of
certain quantities does not demand increasing the size of the supermatrices,
which allows to compute them explicitly.

In all such cases the free energy functional $F\left[ Q\right] $ describing
fluctuations of the $8\times 8$ supermatrices $Q$ can be written as 
\begin{equation}
F[Q]=\frac{\pi \nu }{8}{\rm Str}\int \left[ D\left( \nabla _{{\bf r}}Q({\bf r%
})-\frac{ie}{\hbar c}{\bf A}[Q({\bf r}),\tau _{3}]\right) ^{2}+2i\omega
\Lambda Q({\bf r})\right] {\rm d}{\bf r}\,.  \label{11}
\end{equation}
The $\sigma $-model, Eq.~(\ref{11}), appears after averaging over disorder
and we are to carry out computations with the regular model. In Eq.~(\ref{11}%
) $D$ is the classical diffusion coefficient, $\nu $ is the density of
states, and $\omega $ is the external frequency. The supermatrix $Q$ obeys
the constraint $Q^{2}=1$, ${\bf A}$ is the vector potential corresponding to
an external magnetic field $H$, and the standard notations for the
supertrace ${\rm Str}$ and matrices $\Lambda $, $\tau _{3}$ are used. \cite
{book} Any quantity of interest in the present formulation is expressed as a
functional integral over $Q({\bf r})$ with the weight $\exp (-F[Q])$ and a
combination of elements of the supermatrix $Q$ in the pre-exponential.

We study properties of wave functions calculating their moments written in
Eq.~(\ref{00}). As the first step, we express the moments of the wave
functions in terms of retarded $G^{R}$ and advanced $G^{A}$ Green functions
that can be written as 
\begin{equation}
G_{\varepsilon }^{R,A}\left( x,x^{\prime }\right) =\sum_{\alpha }\frac{\psi
_{\alpha }\left( x\right) \psi _{\alpha }^{\ast }\left( x^{\prime }\right) }{%
\varepsilon -\varepsilon _{\alpha }}\;,  \label{a1}
\end{equation}
where $\psi _{a}$ and $\varepsilon _{\alpha }$ are eigenfunctions and
eigenenergies of the electron states in the disordered system. Using the
spectral expansion, Eq.~(\ref{a1}), assuming that all states are localized,
such that the spectrum is discrete, and considering the most divergent terms
at vanishing frequency $\omega \rightarrow 0$ one can come to the following
relation 
\begin{equation}
\left[ G_{\varepsilon }^{R}(x,x^{\prime })G_{\omega +\varepsilon
}^{A}(x^{\prime },x)\right] ^{n}=\frac{2\pi i(2n-2)!(-1)^{n-1}}{\left(
(n-1)!\right) ^{2}\omega ^{2n-1}}\sum_{\alpha }|\psi _{\alpha
}(x)|^{2n}|\psi _{\alpha }(x^{\prime })|^{2n}\delta (\varepsilon
-\varepsilon _{\alpha })  \label{a2}
\end{equation}

Although Eq.~(\ref{a2}) contains an arbitrary power of the Green functions
we can express the LHS in terms of $8$-component supervectors. This
possibility is due to the fact that the Green functions are taken at two
different points only. The next step is the averaging over disorder and the
decoupling the effective ``interaction'' by integration over the
supermatrices $Q$. A functional integral over $Q$ is simplified in the
saddle-point approximation and we come after standard manipulations \cite
{book} to the following expression 
\begin{equation}
\left\langle \left[ G_{\omega ^{\prime }}^{R}(x,x^{\prime })G_{\omega
+\omega ^{\prime }}^{A}(x^{\prime },x)\right] ^{n}\right\rangle =(\pi \nu
)^{2n}\left\langle [Q_{33}^{12}(x)]^{n}[Q_{33}^{21}(x^{\prime
})]^{n}\right\rangle _{F}\;.  \label{111}
\end{equation}
In the LHS and RHS of Eq.~(\ref{111}), averaging over the impurity potential
and averaging over the free energy, Eq.~(\ref{11}), is implied,
respectively. The superscripts and subscripts of the supermatrices $Q$ stand
for certain matrix elements.

In Section V, exact expressions for the moments of the density-density
correlation function, Eq.~(\ref{00}), as well as its entire distribution
function 
\begin{equation}
{\cal P}_{\psi }(t)=\left\langle \sum_{\alpha }\delta \left( t-\left| \psi
_{\alpha }\left( x\right) \right| ^{2}\left| \psi _{\alpha }\left( x^{\prime
}\right) \right| ^{2}\right) \delta \left( \varepsilon -\varepsilon _{\alpha
}\right) \right\rangle \,.  \label{12}
\end{equation}
will be obtained.

Hereinafter we use dimensionless expression for the wave functions $\psi
SL_{c}\rightarrow \psi $, where $S$ is the cross-section of the wire and $%
L_{c}$ is the localization length. To calculate the distribution function $%
{\cal P}_{\psi }(t)$, Eq.~(\ref{12}), we introduce an auxiliary correlation
function 
\begin{equation}
{\cal P}_{a,b}(t_{1},t_{2})=\left\langle \delta \left(
t_{1}-aQ_{33}^{12}(x_{1})\right) \delta \left(
t_{2}-bQ_{33}^{21}(x_{2})\right) \right\rangle _{F}\,,  \label{13}
\end{equation}
where $a$ and $b$ are some parameters. Once the correlator Eq.~(\ref{13}) is
known, any other quantity of interest can be found by integrations of the
type $\int_{-\infty }^{\infty }{\cal P}_{a,b(a)}\,da$ using a proper choice
of $a$, $b(a)$. As an example, we will calculate the function ${\cal P}%
_{\psi }$ and the distribution 
\begin{equation}
{\cal P}_{Q}=\left\langle \delta \left( t+(i\omega L_{c}\nu
S)^{2}Q_{33}^{12}(x)Q_{33}^{21}(x^{\prime })\right) \right\rangle _{F}\,,
\label{14}
\end{equation}
characterizing the Landauer-type conductivity.\cite{EfId} Hereinafter, all
the correlators are studied in the limit $\omega \rightarrow 0$.

To express the distribution Eq.~(\ref{12}) in terms of a functional integral
over the supermatrices $Q$, we relate the numerical coefficients of all
moments of the distribution functions ${\cal P}_{\psi }$ and ${\cal P}_{Q}$,
using the identity 
\begin{equation}
\frac{(2n-2)!(-1)^{n}}{\left( (n-1)!\right) ^{2}}=\left. \frac{{\rm d}}{{\rm %
d}\beta }\right| _{\beta =1}\int_{0}^{1}{\rm d}p\,(\beta p)^{n-1}\,(\beta
(1-p))^{n-1}\;.  \label{a3}
\end{equation}
Thus, we reduce the entire distribution ${\cal P}_{\psi }$ to ${\cal P}_{Q}$ 
\begin{equation}
{\cal P}_{\psi }(t)=\left. \frac{{\rm d}}{{\rm d}\beta }\right| _{\beta
=1}\int_{0}^{1}\frac{{\rm d}p}{\beta p(1-p)}\left\langle \delta (t+(i\omega
L_{c}S)^{2}\beta ^{2}p(1-p)\,Q_{33}^{12}(x)\,Q_{33}^{21}(x^{\prime
}))\right\rangle _{F}\;  \label{10b}
\end{equation}
Eq.~(\ref{10b}) can be proven taking ${\cal P}_{\psi }\left( t\right) $ from
Eq.~(\ref{12}) and expanding both sides in $\psi _{\alpha }$ and $Q$
respectively. Then, one should use Eqs.~(\ref{a2}, \ref{111}) and this gives
finally Eq.~(\ref{10b}) with the functional $F$ from Eq.~(\ref{11}). Eq.~(%
\ref{10b}) allows us to calculate the distribution function of the
amplitudes of the wave functions at two different points of an infinite
sample.

Although we calculate here the distribution functions of the quantities
written at two different points $x$ and $x^{\prime }$, our procedure can be
extended to many-point functions.

In quasi-1D case, one can calculate the functional integral, Eq.~(\ref{13}),
exactly using the transfer-matrix technique. Within this method the
functional integration is reduced to solving an effective ``Schr\"{o}dinger
equation'' in the space of variables parametrizing $Q$-matrices. \cite{EL}
Solving the ``Schr\"{o}dinger equation'' and calculating certain ``matrix
elements'' that are definite integrals over $Q$ one can obtain any physical
quantity of interest. For instance, the moments corresponding to the
distribution function ${\cal P}_{a,b}$, Eq.~(\ref{13}), read 
\begin{equation}
T_{mn}=\left\langle t_{1}^{n}t_{2}^{m}\right\rangle _{{\cal P}%
_{Q}}=a^{n}b^{m}\int \Psi (Q)\left( Q_{33}^{12}\right) ^{n}\Gamma
(x,x^{\prime };Q,Q^{\prime })\left( Q_{\ 33}^{\prime \,21}\right) ^{m}\Psi
(Q^{\prime })\;{\rm d}Q{\rm d}Q^{\prime }\,,  \label{15}
\end{equation}
where the function $\Psi \left( Q\right) $ is the partition functions of the
parts of the wire to the left from the point $x$ and to the right from the
point $x^{\prime }$, and the function $\Gamma $ represents the partition
functions of the segment between $x$ and $x^{\prime }$.

The meaning of the functions $\Psi $ and $\Gamma $ becomes clear after the
reduction of the functional integrals to the effective Schr\"{o}dinger
equation. \cite{efetov,book,EL,Zirn} The function $\Psi \left( Q\right) $
appears to be the ground state of the effective Hamiltonian ${\cal H}_{Q}$
acting in the space of the elements of the $Q$-matrix whereas the function $%
\Gamma $ is the Green function in this space. They satisfy the following
equations 
\begin{eqnarray}
{\cal H}_{Q}\Psi \left( Q\right) &=&0  \label{a8} \\
\left( \frac{\partial }{\partial x}+{\cal H}_{Q}\right) \Gamma \left(
x,x^{\prime };Q,Q^{\prime }\right) dx &=&\delta \left( x-x^{\prime }\right) 
\nonumber
\end{eqnarray}
The fact that the energy of the ground state is zero is a consequence of the
supersymmetry of the initial $\sigma $-model, Eq.~(\ref{11}).

The explicit form of the Hamiltonian ${\cal H}_{Q}$ depends on the
parametrization of the $Q$-matrix and its choice depends on the quantities
calculated and the physics involved.

The supermatrices $Q$ consist of compact and non-compact blocks containing
circular and hyperbolic functions, respectively. \cite{book} In the cases of
zero and strong magnetic fields only the non-compact sector survives in the
limit $\omega \rightarrow 0$, which greatly simplifies the whole analysis.

If a non-compact variable $\lambda _{1}$ is relevant only, the function $%
\Psi $ takes in the limit of small $\omega $ the well-known form \cite
{Ber,EL,book} 
\begin{equation}
\Psi =2z^{1/2}K_{1}(2z^{1/2}),\text{ }z=i\omega \lambda _{1}  \label{a9}
\end{equation}
valid both for chains and thick wires. Using the modified Bessel functions $%
K_{i\rho }$ one can write in this limit the solution for the function $%
\Gamma $ as 
\begin{equation}
\Gamma (x,x^{\prime };z,z^{\prime })=\frac{8}{\pi ^{2}}\int_{0}^{\infty }%
{\rm e}^{-|x-x^{\prime }|(1+\rho ^{2})/4L_{c}}\;z^{1/2}K_{i\rho
}(2z^{1/2})\;z^{\prime \ 1/2}K_{i\rho }(2z^{\prime \ 1/2})\;\sinh (\pi \rho
)\rho \,{\rm d}\rho \,.  \label{16}
\end{equation}
where $L_{c}$ is the localization length. Eq. (\ref{16}) will be proven in
Sec. V. Actually, it follows from the orthonormality of the modified Bessel
functions $K_{i\rho }$ (see Eq.~(\ref{52})) and the fact that the functions $%
2z^{1/2}K_{i\rho }(2z^{1/2})$ are eigenfunctions of the ``heat'' equation
with the Laplacian in the space of the matrices $Q$. \cite{efetov,book,Zirn}

The expressions for $\Psi $ and $\Gamma $, Eqs.~(\ref{a9}, \ref{16}), are
sufficient to calculate all the moments, Eq.~(\ref{15}), for the orthogonal
and unitary ensembles. This way is more straightforward and more economical,
as compared to the usual way \cite{EL,book} of representing the moments as
solutions of differential equations obtained within the transfer matrix
technique. At the same time, the conventional approach is more general and
is applicable also in the crossover regime.

It is worth mentioning that our analysis of the localization properties in
infinite wires is somewhat easier than that for conductance \cite{Zirn,Ziall}
and wave functions \cite{Fyo-Mir} in finite wires. This is because only
large values of the non-compact variables $\lambda _{1}$ are important for
the calculations in our case. This is this simplification that allows us to
calculate all moments as well as the entire distribution functions for the
`pure' (orthogonal and unitary) ensembles (see Sec. V).

\section{Transfer-Matrix Technique}

At not very high frequencies, the zero transversal space harmonics of the
supermatrix $Q$ in Eq. (\ref{11}) gives the main contribution. In order to
neglect the non-zero harmonics one should choose the London gauge for the
vector potential ${\bf A}$. Writing ${\bf A}=\left( Hy,0,0\right) ,$ where $%
H $ is the external magnetic field perpendicular to the wire we represent
the free energy, Eq.~(\ref{11}), for the quasi-1D sample in the form 
\begin{equation}
F[Q]=\frac{L_{{\rm cu}}}{16}{\rm Str}\int \left[ \left( \nabla _{x}Q\right)
^{2}+\frac{X^{2}}{16L_{{\rm cu}}^{2}}[Q,\tau _{3}]^{2}+\frac{2i\omega
\Lambda Q(x)}{D}\right] {\rm d}x\,,  \label{21}
\end{equation}
where $L_{{\rm cu}}=2\pi S\nu D$ is the localization length for the unitary
ensemble. The crossover parameter $X$ depends on the geometry of the sample, 
$X=2\pi \phi /\phi _{0}$, $\phi _{0}=hc/e$ is the flux quantum, and $\phi
=HL_{{\rm cu}}\left\langle y^{2}\right\rangle _{\sec }^{1/2}$ is the
magnetic flux through the area limited by the localization length. The
brackets imply averaging across the wire over the coordinate $y$ which is
chosen to be perpendicular both to the direction of the wire and the
magnetic field. This averaging gives $\left\langle y^{2}\right\rangle
^{1/2}=d/\sqrt{12}$, where $d$ is the width of the wire, for ``flat'' wires
made on the basis of a 2D gas. \cite{Exp1,Exp2} For wires with a circular
cross-section, $\left\langle y^{2}\right\rangle ^{1/2}=d/4$, where $d$ is
the diameter.

To determine the functions $\Psi $ and $\Gamma $ in Eq.~(\ref{15}), we
discretize the wire by introducing sites on which the $Q$-supermatrix
varies. The function $\Psi \left( Q\right) $ is the partition function of a
semi-infinite sample and therefore it does not depend on the coordinate $x$.
This allows us to write immediately the a recurrence equation. Relating the
function $\Psi (Q)$ on one site and its value $\Psi \left( Q^{\prime
}\right) $ on the neighboring one we obtain 
\begin{equation}
\Psi (Q)=\int N(Q,Q^{\prime })Z_{0}(Q^{\prime })\Psi (Q^{\prime }){\rm d}%
\,Q^{\prime }\,.  \label{22}
\end{equation}
where the function $N\left( Q,Q^{\prime }\right) $ describing the coupling
between the sites 
\begin{equation}
N(Q,Q^{\prime })=\exp \left( \frac{L_{{\rm cu}}}{4\Delta l}{\rm Str}%
\;QQ^{\prime }\right) \,,  \label{23}
\end{equation}
originates from the kinetic term $(\nabla Q)^{2}$ in the free energy, Eq.~(%
\ref{21}). The function $Z_{0}\left( Q\right) $, 
\begin{equation}
Z_{0}(Q)=\exp \left( \frac{i\omega L_{{\rm cu}}\Delta l}{4D}{\rm Str}%
\,\Lambda Q+\frac{X^{2}\Delta l}{16L_{{\rm cu}}}{\rm Str}\,[Q,\tau
_{3}]^{2}\right)  \label{24}
\end{equation}
describes the remaining terms in the free energy functional $F$, Eq. (\ref
{21}), and should be integrated on each site. In order to reduce the
integral Eq. (\ref{22}) to the differential ``Schr\"{o}dinger equation'' for
the functions $\Psi $ (and similarly for $\Gamma $), one should take the
continuous limit $\Delta l\rightarrow 0$, where $\Delta l$ is the length of
the separation between the sites. In this continuous limit, $\Delta l$ drops
out from final expressions and therefore can be chosen to be arbitrarily
short.

Since the kernel $\Gamma $ enters equations for each moment only in a
combination with $(Q^{21})^{n}\Psi (Q)$, it is advantageous to introduce a
matrix function $P_{k}^{(n)}$ such that 
\begin{equation}
P_{k}^{(n)}(Q)=\sum_{x}\exp [i(x-x^{\prime })k]\int \Gamma (k;Q,Q^{\prime
})(Q^{\prime \,21})^{n}\Psi (Q^{\prime })\,{\rm d}Q^{\prime }\,,  \label{25}
\end{equation}
where $\Gamma (k;Q,Q^{\prime })$ is the Fourier transform of $\Gamma
(x-x^{\prime };Q,Q^{\prime })$. The matrix function $P_{k}^{(n_{)}}(Q)$ on a
site is related to its value $P_{k}^{(n_{)}}(Q^{\prime })$ on the
neighboring site as 
\begin{equation}
P_{k}^{(n)}(Q)-\exp (-ik\Delta l)\int N(Q,Q^{\prime })Z_{0}(Q^{\prime
})P_{k}^{(n)}(Q^{\prime }){\rm d}Q^{\prime }=(Q^{21})^{n}\Psi (Q)\,.
\label{26}
\end{equation}
The moments of the distribution ${\cal P}_{Q}$, expressed in terms of the
functions $\Psi $ and $P_{k}$, are given by 
\begin{equation}
T_{mn}=\int \Psi (Q)(Q_{33}^{21})^{n}(P_{k\,33}^{(m)}+P_{-k\,33}^{(m)}){\rm d%
}Q\,,  \label{27a}
\end{equation}
Solving Eqs. (\ref{25}-\ref{26}) and calculating the integral in Eq. (\ref
{27a}) one can find, in principle, all the moments. However, further
simplifications are necessary to obtain the results explicitly.

A spectral expansion \cite{efetov,book,Zirn} is convenient for the analysis
of the function $P_{k}^{(n)}$. So, we expand the functions $P_{k}^{(n)}$ in
eigenfunctions $\phi _{{\cal E}}(Q)$, introduced as 
\begin{equation}
\int N(Q,Q^{\prime })Z_{0}(Q^{\prime })\phi _{{\cal E}}(Q^{\prime }){\rm d}%
Q^{\prime }={\cal E}\phi _{{\cal E}}(Q)\,.  \label{27}
\end{equation}
The matrix functions $\phi _{{\cal E}}(Q)$ have the same structure as $%
Q^{21} $. Their orthogonality and the normalization properties can be
written in the form 
\begin{equation}
\int [\phi _{{\cal E}^{\prime }}^{+}(Q)]_{33}[\phi _{{\cal E}}(Q)]_{33}{\rm d%
}Q=\delta ({\cal E}-{\cal E}^{\prime })\,.  \label{28}
\end{equation}
Expanding $P_{k}^{(m)}$ in the complete set of the functions $\phi _{{\cal E}%
}$, Fourier transforming it to the coordinate representation, and
substituting the result into Eq.~(\ref{15}), we arrive at 
\begin{equation}
T_{mn}=-\sum_{{\cal E}}\exp (-{\cal E}|x-x^{\prime }|)\int
(Q_{33}^{12})^{m}[\phi _{{\cal E}}(Q)]_{33}\Psi (Q)\,{\rm d}Q\int
(Q_{33}^{21})^{n}[\phi _{{\cal E}}(Q)]_{33}\Psi (Q)\,{\rm d}Q\,.  \label{29}
\end{equation}
Eq. (\ref{29}) contains the sum over all eigenstates that should be found
from Eq. (\ref{27}) and establishes a link between the eigenvalue problem in
the space of parameters of the $Q$-matrix and the localization properties.
The lowest non-zero eigenenergy corresponding to the first excited state
determines the exponential decay of the correlation functions at large
distances.

\section{Effective Hamiltonians}

Eqs. (\ref{22}, \ref{27}, \ref{29}) form a closed system completely solving
the problem. To perform explicit calculations a parametrization of the $Q$%
-matrices should be chosen. For the problem involved one can use either the
``standard''\cite{efetov,book} or the ``magnetic'' parametrization\cite
{EfAl,book}. The former parametrization has been used, e.g., for calculation
of the density-density correlation function in disordered wires for the
orthogonal and unitary ensembles, while the latter one helped to obtain
explicit formulae for the crossover between the ensembles in the
zero-dimensional situation.

For the problem of the localization in wires the standard parametrization
allows to calculate any quantity of interest for the orthogonal and unitary
ensembles and it will be used in Sec. V for calculations of all moments as
well as of the distribution function of the density-density correlator. At
the same time, the standard parametrization is of little help in the
crossover regime between the ensembles and we will use in this case the
magnetic parametrization.

The basic equations and their solutions for the orthogonal and unitary
ensembles can be found in Refs. \cite{efetov,book} and we do not write them
here. Instead, let us present explicit differential equations that can be
obtained within the magnetic parametrization from Eqs. (\ref{22}, \ref{27}, 
\ref{29}).

The $Q$-matrix in this parametrization is constructed from two sets of
variables, cooperon ones being cut by the magnetic field, and diffuson ones
insensitive to $H$. An explicit form of the supermatrices $Q$ in these two
parametrizations can be found elsewhere. \cite{book}The supermatrix $Q$ in
this parametrization is represented as 
\begin{equation}
Q=V_{d}Q_{c}\bar{V}_{d}\text{, }  \label{e1}
\end{equation}
where the supermatrix $Q_{c}$, $Q_{c}^{2}=1$, contains cooperon ``degrees of
freedom'' and $V_{d}$, $\bar{V}_{d}V_{d}=1$, diffuson ones. To make the
reading easier, we present here the essential structure of the block $Q^{12}$
in the magnetic parametrization, keeping only those terms that contribute to
the correlators considered 
\begin{equation}
\begin{array}{c}
Q^{12}=u_{d}i(1-\hat{\lambda}_{d}^{2})^{1/2}[\hat{\lambda}_{c}+2(\eta
_{c}\eta _{c}^{\ast }-\kappa _{c}\kappa _{c}^{\ast })(\lambda _{1c}-\lambda
_{c})]\overline{v}_{d}\,,
\end{array}
\label{100}
\end{equation}
where $\hat{\lambda}_{c,\,d}={\rm diag}(\cos \theta _{c,\,d},\cosh \theta
_{1c,\,1d})$, $v_{d}$ and $u_{d}$ are the standard $4\times 4$ unitary
matrices in the $V_{d}$ (diffuson) block, and $\eta _{c}$, $\eta _{c}^{\ast
} $, $\kappa _{c}$, $\kappa _{c}^{\ast }$ are the Grassmann variables from
the cooperon block.

An advantageous feature of the magnetic parametrization is that the term
containing the magnetic field $H$ in the free energy has a simple form 
\begin{equation}
{\rm Str}[Q,\tau _{3}]^{2}=16(\lambda _{1c}^{2}-\lambda _{c}^{2})\,.
\label{32}
\end{equation}
In contrast, the term with the frequency $\omega $ is quite complex since it
contains the Grassmann variables 
\begin{equation}
{\rm Str}(\Lambda Q)=4[\lambda _{1d}\lambda _{1c}-\lambda _{d}\lambda
_{c}+2(\eta _{c}\eta _{c}^{\ast }-\kappa _{c}\kappa _{c}^{\ast })(\lambda
_{1c}-\lambda _{c})(\lambda _{1d}-\lambda _{d})]\;.  \label{33}
\end{equation}

The functions $\Psi $ and $\phi _{{\cal E}}$ correspond to states having
respectively the ``angular momentum'' zero and one with respect to the
unitary rotations $V_{d}$. The form of the functions $\phi _{{\cal E}}$ is
determined by the RHS of Eq. (\ref{26}) and, as a result, it should be
searched for in the class of functions with the same structure as the matrix 
$Q^{12}$

\begin{equation}
\phi _{{\cal E}}=v_{d}\left( 
\begin{array}{cc}
(1-\lambda _{d}^{2})^{1/2}f & 0 \\ 
0 & i(\lambda _{1d}^{2}-1)^{1/2}f_{1}
\end{array}
\right) \bar{u}_{d}\,.  \label{34}
\end{equation}
where $v_{d}$ and $u_{d}$ are unitary matrices containing Grassmann
variables only. The different forms of the functions $\Psi $ and $\phi _{%
{\cal E}\text{ }}$ lead to different forms of the effective Hamiltonians
determining the functions $\Psi $ and $\phi _{{\cal E}}$. The functions $f$
and $f_{1}$ do not contain the diffuson Grassmann variables but may contain
all other variables.

The Schr\"{o}dinger equations corresponding to Eqs.~(\ref{22}) and (\ref{27}%
) are obtained considering slow variations of $Q$ on neighboring sites. The
matrix $Q^{\prime }$ is expanded near $Q$ and integrated over small
variations of the matrix elements. The expansion up to quadratic variations
is sufficient and therefore, the quadratic form obtained is identical to the
one describing the square of the elementary length in the space of the
supermatrices $Q$. \cite{book} This elementary length determines the
Berezinian corresponding to a chosen parametrization. Correspondingly, the
``Laplacian'' entering the effective Schr\"{o}dinger equation contains the
Berezinian of the transformation. The Berezinian of the magnetic
parametrization reads \cite{EfAl,book} 
\begin{equation}
J=J_{c}J_{d}J_{cd},\text{ }\,J_{c,d}=\frac{1}{2^{6}\pi ^{2}(\lambda
_{1c,d}-\lambda _{c,d})^{2}}\,,\text{ }\,J_{cd}=\frac{4\lambda _{c}^{2}}{%
(\lambda _{1c}+\lambda _{c})^{2}},  \label{35}
\end{equation}
The parts $J_{d}$ and $J_{cd}$ of the Berezinian Eq.~(\ref{35}) originate
from the diffuson Grassmann variables $\eta _{d}$, $\kappa _{d}$, whereas $%
J_{c}$ comes from the cooperon ones $\eta _{c}$, $\kappa _{c}$. The diffuson
Grassmann variables do not enter Eqs.~(\ref{32}) and (\ref{33}). In some
sense, the diffuson variables decouple from other variables and this
simplifies calculations.

The situation with the cooperon variables is more difficult. Since the free
energy, Eqs.~(\ref{21}, \ref{33}), contains explicitly the cooperon
Grassmann variables $\eta _{c}$ and $\kappa _{c}$, so do the functions $\Psi 
$ and $\phi _{{\cal E}}$. In order to write equations for these functions
explicitly one should expand them in powers of the Grassmann variables.
Proceeding in this way one obtains a system of differential equations for
the functions entering the expansion. Another complication occurs due to a
non-trivial angular dependence on $Q_{c}$. This matrix is surrounded by the
diffuson modes and couples to them. As a result, one comes to extremely
cumbersome equations that can hardly be solved analytically for arbitrary $X$%
. The solution can be found for $X\gg 1$ but this limit corresponds to the
well studied unitary ensemble.

Fortunately, an important information about correlation functions can be
extracted in the limit $X\ll 1$. To understand why this limit helps to study
the problem we remind the reader that the possibility to obtain closed
expressions for correlation functions in the orthogonal and unitary
ensembles is due to the fact that the main contribution to integrals comes
in the limit of small frequencies $\omega $ from large values of the
non-compact variables (in the unitary ensemble $\lambda _{1d}\sim 1/\omega $%
). For large values of $\lambda _{1d\text{ }}$ the partial differential
equations for the functions $\Psi $ and $\phi _{{\cal E}}$ are sufficiently
simple and can be solved. In the crossover regime when $X$ is finite, the
equations are still very complicated even in the limit $\omega \rightarrow 0$%
. However, in the limit $X\ll 1$, some quantities of interest are determined
by large $\lambda _{1c}\sim 1/X\gg 1$. This leads to an additional
simplification of the equations and the possibility to make estimates. At
the same time, not any correlation function can be calculated in this way,
which limits the applicability of the trick.

In the continuous limit, Eq. (\ref{22}) can be reduced to the form 
\begin{equation}
{\cal H}_{X}\Psi =0  \label{36}
\end{equation}

Due to the presence of the Grassmann variables in the free energy
functional, Eqs. (\ref{21}, \ref{33}), the solution $\Psi $ must contain
them too, which is in contrast to the solutions for the orthogonal and
unitary ensembles.\ The function $\Psi $ can be represented as 
\begin{equation}
\Psi =\Psi _{0}\left( \lambda _{1d},\lambda _{1c}\right) +\Psi _{1}\left(
\lambda _{1c},\lambda _{1d}\right) \left( \eta _{c}\eta _{c}^{\ast }-\kappa
_{c}\kappa _{c}^{\ast }\right) +\Psi _{2}\left( \lambda _{1c},\lambda
_{1d}\right) \eta _{c}\eta _{c}^{\ast }\kappa _{c}\kappa _{c}^{\ast }
\label{38a}
\end{equation}
Then, one obtains a system of partial differential equations for the
functions $\Psi _{0}$, $\Psi _{1}$, $\Psi _{2}$%
\[
\left( {\cal H}_{X}^{(0)}+\tilde{\omega}x_{1}\right) \Psi _{0}+2\frac{%
\lambda _{1c}\lambda _{c}-1}{\left( \lambda _{1c}-\lambda _{c}\right) ^{2}}%
\Psi _{1}=0 
\]
\begin{equation}
\left( {\cal H}_{X}^{(0)}+\tilde{\omega}x_{1}\right) \Psi _{1}-\frac{\lambda
_{1c}\lambda _{c}-1}{\left( \lambda _{1c}-\lambda _{c}\right) ^{2}}\Psi _{2}+%
\tilde{\omega}x_{2}\Psi _{0}=0  \label{38b}
\end{equation}

\[
\left( {\cal H}_{X}^{(0)}+\tilde{\omega}x_{1}\right) \Psi _{2}-2\tilde{\omega%
}x_{2}\Psi _{1}=0 
\]
where 
\begin{equation}
{\cal H}_{X}^{(0)}=-\sum_{i=d,1d}\frac{1}{J_{d}}\frac{\partial \,\,\,}{%
\partial \lambda _{i}}J_{d}\,|1-\lambda _{i}^{2}|\frac{\partial \,\,\,}{%
\partial \lambda _{i}}-\sum_{i=c,1c}\frac{1}{J_{c}J_{cd}}\frac{\partial
\,\,\,}{\partial \lambda _{i}}J_{c}J_{cd}\,|\lambda _{i}^{2}-1|\frac{%
\partial \,\,\,}{\partial \lambda _{i}}+X^{2}\left( \lambda
_{1c}^{2}-\lambda _{c}^{2}\right) \,.  \label{38c}
\end{equation}
\[
\tilde{\omega}=2\pi ^{2}\left( \nu S\right) ^{2}D\omega 
\]
and

\begin{equation}
x_{1}=\lambda _{1d}\lambda _{1c}-\lambda _{d}\lambda _{c},\text{ \ }%
x_{2}=2\left( \lambda _{1c}-\lambda _{c}\right) \left( \lambda _{1d}-\lambda
_{d}\right)  \label{38d}
\end{equation}

In the limit $\lambda _{1d}$, $\lambda _{1c}\gg 1,$ the difference between
equations for the functions $\Psi $ and $\phi _{{\cal E}}$ is negligible
because it comes from the averages of the type $\langle \Delta \eta
_{i}\Delta \eta _{i}^{\ast }\rangle \sim 1/\lambda _{1i}$, $\langle \Delta
\kappa _{i}\Delta \kappa _{i}^{\ast }\rangle \sim 1/\lambda _{1i}$, $i=c$, $%
d $. In this limit, Eqs. (\ref{38b}, \ref{38c}, \ref{38d}) take a simpler
form 
\[
\left( {\cal H}_{X}^{(0)}+\tilde{\omega}x_{1}\right) \Psi _{0}=0 
\]
\begin{equation}
\left( {\cal H}_{X}^{(0)}+\tilde{\omega}x_{1}\right) \Psi _{1}+\tilde{\omega}%
x_{2}\Psi _{0}=0  \label{e38}
\end{equation}

\[
\left( {\cal H}_{X}^{(0)}+\tilde{\omega}x_{1}\right) \Psi _{2}-2\tilde{\omega%
}x_{2}\Psi _{1}=0 
\]
with 
\begin{equation}
{\cal H}_{X}^{(0)}=-\lambda _{1d}^{2}\frac{\partial ^{2}\,\,\,}{\partial
\lambda _{1d}^{2}}-\lambda _{1c}^{2}\frac{\partial ^{2}\,\,\,\,}{\partial
\lambda _{1c}^{2}}+2\lambda _{1c}\frac{\partial \,\,\,\,}{\partial \lambda
_{1c}}+X^{2}\lambda _{1c}^{2}  \label{37}
\end{equation}

\[
x_{1}=x_{2}/2=\lambda _{1c}\lambda _{1d} 
\]
The same equations can be written in the limit $\lambda _{1d}$, $\lambda
_{1d}\gg 1$ for the functions $\phi _{{\cal E}}$.

It is clear that Eqs. (\ref{38a}, \ref{38d}) cannot be solved analytically
for an arbitrary $X$ and the only hope is to solve Eqs. (\ref{e38}, \ref{37}%
). Of course, one is restricted then by calculation of those quantities for
which large $\lambda _{1d},\lambda _{1c}$ give the main contribution.

In the opposite limiting case $X\rightarrow \infty $ but $\omega \rightarrow
0$ all the cooperon variables are frozen and one comes to the effective
Hamiltonian 
\begin{equation}
{\cal H}=-\lambda _{1}^{2}\frac{\partial ^{2}\,\,\,}{\partial \lambda
_{1}^{2}}-\frac{2i\omega L_{{\rm cu}}^{2}}{D}\lambda _{1}  \label{39a}
\end{equation}

After a proper change of the localization length the same Hamiltonian can be
written in the orthogonal case using the standard parametrization. For
completeness let us remind that the localization lengths for the orthogonal
and unitary ensembles, $L_{{\rm co}}$ and $L_{{\rm cu}}$ respectively, are
given by the following expressions \cite{EL,book}

\begin{equation}
\begin{array}{cc}
L_{{\rm cu}}=2\pi \nu SD\;, & L_{{\rm co}}=\pi \nu SD\;,
\end{array}
\label{Lci}
\end{equation}

In these cases the function $\Psi $ does not contain Grassmann variables.
The equations to be solved take the form 
\begin{equation}
{\cal H}\,\Psi =0\,,\ {\cal H}\,\phi _{{\cal E}}=L_{c}{\cal E}\,\phi _{{\cal %
E}}  \label{40a}
\end{equation}
$\;$where the localization length $L_{c}$ equals either $L_{{\rm co}}$ or $%
L_{{\rm cu}}$ depending on the ensemble.

In the next Section, solving Eqs. (\ref{40a}) we will calculate all moments
and the entire distribution function of the density-density correlations.
This will help us to come to certain conclusions about properties of the
electron wave functions in the orthogonal and unitary ensembles.

\section{Calculation of Moments and Distributions for the Orthogonal and
Unitary Ensembles.}

Calculation of the moments and of the distribution functions is important
for the understanding of the structure of the wave functions and properties
of different averages. Higher moments and the distribution function for
infinitely long wires have not been considered yet although the results are
known for finite wires. \cite{Fyo-Mir} So, calculation of the moments and of
the distribution function can be interesting on its own but, what is more
important, this information will also help us to make certain conclusions
about properties of the wave functions in a weak magnetic field where the
possibility of exact calculations is more limited. This will help us to
extend the results of Ref. \cite{WE} where only the averaged density-density
correlation function was calculated.

Solutions $\Psi $ and $\phi _{{\cal E}}$ of Eqs. (\ref{39a}, \ref{40a}) are
well known (see e.g. \cite{efetov,book,Zirn}) and can be written as 
\begin{equation}
\begin{array}{ccc}
\Psi =2z^{1/2}K_{1}(2z^{1/2})\,, & \phi _{{\cal E}}=2z^{1/2}K_{i\rho
}(2z^{1/2})\,, & {\cal E}=(1+\rho ^{2})/4L_{c}\;,
\end{array}
\label{51}
\end{equation}
where $z=2i\lambda _{1}\omega L_{c}^{2}/D$. Comparing Eqs.~(\ref{29}) and (%
\ref{15}), we can understand that the propagator has the form of Eq.~(\ref
{16}). The numerical coefficient in Eq. (\ref{16}) is found using the
Lebedev-Kontorovich transformation \cite{manual,AltPrig} applicable for an
arbitrary function $\varphi (y)$ 
\begin{equation}
\begin{array}{cc}
\widetilde{\varphi }(\rho )=\int_{0}^{\infty }\varphi (y)K_{i\rho }(y)\frac{%
{\rm d}y}{y}\,, & \varphi (y)=\frac{2}{\pi ^{2}y}\int_{0}^{\infty }%
\widetilde{\varphi }(\rho )K_{i\rho }(y)\rho \sinh (\pi \rho ){\rm d}\rho .
\end{array}
\label{52}
\end{equation}

Now we can calculate all the moments of the two-point correlator, Eq.~(\ref
{15}). Substituting Eqs.~(\ref{51}) into Eq.~(\ref{29}), and calculating
integrals over all elements of the supermatrices $Q$ except $\lambda _{1}$,
yields 
\begin{equation}
T_{mn}=\left\langle t_{1}^{n}t_{2}^{m}\right\rangle _{{\cal P}_{Q}}=\frac{32%
}{\pi ^{2}}\int_{0}^{\infty }{\rm d}\rho \,\rho \sinh (\pi \rho )e^{-\frac{u%
}{4}\left( \rho ^{2}+1\right) }M_{n}\left( \rho \right) M_{m}\left( \rho
\right) \,\;.  \label{53}
\end{equation}
where 
\[
M_{n}=\int {\rm d}t_{1}\left( \frac{t_{1}}{2}\right)
^{2n-1}n^{2}K_{1}(t_{1})K_{i\rho }(t_{1})\;. 
\]
and $u=|x-x^{\prime }|/L_{c}$.

The structure of Eq.~(\ref{53}) is not complicated and the integration over
the variable $t_{1}$ can be performed exactly\cite{GradRyzh}. As a result,
we come to general expressions for all moments of the density-density
correlations $p_{\infty }^{\left( n\right) }\left( u\right) $ introduced in
Eq. (\ref{00}) 
\begin{equation}
p_{\infty }^{(n)}\left( u\right) =\frac{2\nu }{\pi }\frac{(2n-2)!}{\left(
(n-1)!\right) ^{2}}\int_{0}^{\infty }{\rm d}\rho \,\rho \sinh (\pi \rho )%
{\rm e}^{-\frac{u}{4}\left( \rho ^{2}+1\right) }\left[ P^{(n)}(\rho )\right]
^{2}\;,  \label{58}
\end{equation}
\[
P^{(n)}(\rho )=\frac{n^{2}}{(2n-1)!}\Gamma \left( n+\frac{1}{2}+\frac{i\rho 
}{2}\right) \Gamma \left( n+\frac{1}{2}-\frac{i\rho }{2}\right) \Gamma
\left( n-\frac{1}{2}+\frac{i\rho }{2}\right) \Gamma \left( n-\frac{1}{2}-%
\frac{i\rho }{2}\right) \;. 
\]
For $n=1$, Eq.~(\ref{58}) reduces to the expression obtained by Gogolin \cite
{Gogol} for a single chain. Eq. (\ref{58}) is applicable at any $x$
including the values smaller and of the order of $L_{c}$ for both the
unitary and orthogonal ensembles. Comparing the moments, Eqs. (\ref{58}),
with those obtained for finite wires \cite{Fyo-Mir}, we see that their
structure is similar, although they are not completely identical.

At large distances, $|x-x^{\prime }|\gg L_{c}$, essential contribution to
the integral over $\rho $ in Eq. (\ref{58}) comes from small $\rho \ll 1$
and the integral can be easily calculated. We see that the moments of all
quantities decay similarly as in infinite wires, Eqs.~(\ref{58}), (\ref{53}%
), finite closed wires \cite{Fyo-Mir}, as well as in open chains \cite
{1chain1,1chain2} 
\begin{equation}
p_{\infty }^{(n)}(x)=C_{n}\left( \frac{1}{|x|}\right) ^{3/2}\,\exp \left( -%
\frac{|x|}{4L_{c}}\right) \;.  \label{a102}
\end{equation}
where $C_{n}$ is a constant depending on $n$. The exponential decay of all
moments seen from Eq. (\ref{a102}) manifests the localization of the wave
functions. A very important feature of the moments $p_{\infty }^{(n)}(x)$,
Eq. (\ref{a102}), is that the dependence on the coordinate is the same for
all $n$. This does not allow to interpret $p_{\infty }^{\left( n\right)
}\left( x\right) $, Eq. (\ref{a102}), as functions characterizing the
typical shape of the wave functions.

The exact representation of the moments of the wave functions, Eq. (\ref{53}%
), allows to obtain the entire distribution function of the density-density
correlations. Of the crucial importance is the fact that Eq. (\ref{53}) is
valid for all $n$ including the limit $n\rightarrow \infty $. It is clear,
if the $n$th moment of a quantity can be represented in a form of an
integral of a function multiplied by the $n$th power of the variable of
integration, then this function is the distribution of this quantity. In the
case under consideration, Eq.~(\ref{53}) contains, besides the necessary
power of the variable $t$, the factor $n^{2}$, originating from the
expansion of the element $(Q_{33}^{12})^{n}$ in the Grassmann variables.
This does not make the calculation of the distribution function more
difficult because the presence of this factor leads merely to additional
derivatives, and we obtain for the distribution function ${\cal P}%
(t_{1},t_{2})$, Eq. (\ref{13}) 
\begin{equation}
{\cal P}(t_{1},t_{2})=\frac{2}{\pi ^{2}}\int_{0}^{\infty }{\rm d}\rho \,\rho
\sinh (\pi \rho ){\rm e}^{-\frac{u}{4}\left( \rho ^{2}+1\right) }\,\Pi
_{\rho }(t_{1})\,\Pi _{\rho }(t_{2})\;,  \label{54}
\end{equation}
where 
\begin{equation}
\Pi _{\rho }(t)=\left( \left[ K_{1}(2t^{1/2})K_{i\rho }(2t^{1/2})\right]
^{\prime }t\right) ^{\prime }\;.  \label{54a}
\end{equation}
and $^{\prime }$ stands for the derivative over $t$. The two-point
distribution function ${\cal P}(t_{1},t_{2})$ can be used to obtain any
other correlator of interest. For instance, the distribution of the
Landauer-type conductivity ${\cal P}_{Q}(t)$, Eq.~(\ref{14}) can be
represented as the convolution 
\begin{equation}
{\cal P}_{Q}=\int {\rm d}\tau \left\langle \delta (t-\tau i\omega \nu
L_{c}SQ_{33}^{12}(x_{1}))\delta (\tau -i\omega \nu
L_{c}SQ_{33}^{21}(x_{2}))\right\rangle _{F}\;.  \label{a100}
\end{equation}
Then, we obtain for the function ${\cal P}_{Q}\left( t\right) $ 
\begin{equation}
{\cal P}_{Q}(t)=\frac{2\nu }{\pi ^{2}}\int_{0}^{\infty }{\rm d}\rho \,\rho
\sinh (\pi \rho ){\rm e}^{-\frac{u}{4}\left( \rho ^{2}+1\right)
}\,\int_{0}^{\infty }\frac{{\rm d}\tau }{\tau }\Pi _{\rho }(\tau )\,\Pi
_{\rho }(t/\tau )\;,  \label{55}
\end{equation}

Eq. (\ref{55}) is also exact and applicable at any distances. It is
interesting to find this function near its maximum and this can be done
easily in the limit $|x-x^{\prime }|\gg L_{c}$. At very large separation
between $x$ and $x^{\prime }$, typical values of $t$ are exponentially small
in distributions ${\cal P}_{Q}$ and ${\cal P}_{\psi }$, Eqs. (\ref{14}, \ref
{10b}). A dominant contribution to the integral, Eq.~(\ref{55}), comes from
the region $\tau \sim t\ll 1$. Eq.~(\ref{54a}) allows us to write the
asymptotics of the function $\Pi _{\rho }(\tau )$ at $\tau \ll 1$ and $\tau
\gg 1$ 
\[
\Pi _{\rho }(\tau )\simeq 
%TCIMACRO{\func{Re}}%
%BeginExpansion
\mathop{\rm Re}%
%EndExpansion
\left[ \Gamma (i\rho )/4\tau ^{3/2+i\rho /2}\;\right] ,\text{ }\tau \ll 1 
\]
\begin{equation}
\Pi _{\rho }(\tau )\simeq \pi \exp \left( -4\tau ^{1/2}\right) /\tau ^{1/2},%
\text{ }\tau \gg 1  \label{ab103}
\end{equation}
which ensures convergence of the integral over $\tau $ at $\tau \rightarrow
0 $ in Eq. (\ref{55}). Performing integration in the resulting expression 
\[
{\cal P}_{Q}(-\ln t)=\frac{\pi }{32}\int_{-\infty }^{\infty }{\rm d}\rho
\,\rho \sinh (\pi \rho )\frac{\Gamma (i\rho )\Gamma (2+i\rho )}{t^{1/2}}%
\,\exp \left( -\frac{u}{4}-\frac{\ln ^{2}t}{4u}-\frac{u}{4}(\rho +\frac{i\ln
t}{u})^{2}\right) \;, 
\]
in the saddle point approximation, we obtain 
\begin{equation}
{\cal P}_{Q}(-\ln t)\simeq \frac{1}{2\pi ^{1/2}u^{1/2}}\exp \left( -\frac{1}{%
4u}(u+\ln t)^{2}\right) \;.  \label{59}
\end{equation}

Considering the distribution ${\cal P}_{\psi }$, Eq.~(\ref{10b}), under the
same assumptions it is not difficult to check that in the limit $%
|x-x^{\prime }|\gg L_{c}$ it reduces to Eq.~(\ref{59}).

The saddle-point approximation used for the derivation of Eq. (\ref{59}) is
applicable at large distances $|x-x^{\prime }|\gg L_{c}$ when $u$ is large.
Eq. (\ref{59}) shows that in this limit the distribution functions are
log-normal. The result about the log-normal distribution of physical
quantities like conductance agrees with the corresponding result obtained
for chains \cite{1chain1,1chain2} and is universal. It shows that the
localization length of the wave functions or in, other words, the logarithm
of the conductance is a self-averaging quantity. Considering very long
samples one has to obtain the localization length $L_{c}$ and fluctuations
of this quantity can be neglected.

At the same time, the log-normal distribution, Eq. (\ref{59}), is applicable
for not very large values of $t$ only. The far tail of the distribution
function ${\cal P}_{\psi }$ decays more slowly. The moments of the
density-density correlations $p_{\infty }^{\left( n\right) }$, Eq. (\ref
{a102}), cannot be obtained from Eq. (\ref{59}) because the main
contribution comes from the far tails of the distribution function. All
these peculiarities are discussed in details in Refs. \cite{1chain1,1chain2}.

What is the physical meaning of the fact that all the moments $p_{\infty
}^{\left( n\right) }$, Eq. (\ref{a102}) decay in the same way with a larger
length $4L_{c}$? This can be understood if we imagine that there are big
splashes of the wave functions even far from the localization center.
Although, they can give a considerable contribution to the shape of the wave
function, their probability decreases exponentially with $|x|$ (is
proportional to $\exp \left( -|x|/4L_{c}\right) $). Then, such splashes
would give a similar contribution to all the moments. As concerns typical
values of the squares of wave functions, they decay faster with the
localization length $L_{c}$ and give a smaller contribution to the moments.

The situation is opposite if one calculates the average logarithm of the
wave functions. For the logarithm, the splashes do not give a specially
large contribution, being at the same time very rare. So, their contribution
to the average logarithm can be neglected. In other words, the average
logarithm of the correlations of wave functions characterizes typical wave
functions, whereas the moments describe the rare splashes of the wave
functions. This understanding will be important for interpretation of the
results in a weak magnetic fields obtained in the next Section.

\section{Effect of Magnetic Field on the Tails of Wave Functions}

This section is central in our paper. We try now describe electron wave
functions in disordered wires in a weak magnetic field, such that $X\ll 1$.
General equations describing this situation are very difficult even when
they are written for the ``ground state'' function $\Psi $, Eqs. (\ref{36}-%
\ref{38d}). At the same time, any attempt to consider the weak magnetic
field as a perturbation fails in the most interesting limit $\omega
\rightarrow 0$ (one obtains terms like $X^{2}/\tilde{\omega}$). This
reflects a clear physics: travelling a sufficiently long time the electron
`feels' any weak magnetic field.

The failure of the perturbation theory signals immediately that the effect
of the weak magnetic field is more complicated than a slight changing of the
eigenenergies of the effective Sch\"{o}dinger equation and hence the
localization length. The first excited state for the orthogonal ensemble has
the eigenenergy ${\cal E}_{{\rm o}}=1/2$ and this energy definitely changes
if the magnetic field is applied. But, as we will try to show below, an
additional eigenstate with the eigenenergy ${\cal E}_{{\rm u}}=1/4$, the
energy of the unitary ensemble, appears (strictly speaking, a continuum of
states with the energies ${\cal E}_{\rho }=(1+\rho ^{2})/4$, where $\rho $
is a continuous variable, should be added). Although at $X\ll 1$ this state
contributes to correlation functions with a small weight, the value ${\cal E}%
_{{\rm u}}=1/4$ does not depend on $X$.

As concerns the eigenenergy of the ground state, it is exactly zero due to
the supersymmetry of the model. The eigenfunction $\Psi $ of the ground
state is a complicated function $Q$ depending on the ratio $X^{2}/\tilde{%
\omega}.$ We did not manage to find it for an arbitrary value of this ratio
and can write in the limits $X^{2}/\tilde{\omega}\rightarrow 0$ (orthogonal
ensemble) and $X^{2}/\tilde{\omega}\rightarrow \infty $ (the limit we are
interested in now) only. Although the solution for the orthogonal ensemble
is very well known in the standard parametrization \cite{efetov,book}, it
has not been written before in the magnetic parametrization. To find it in
the limit $\lambda _{1d}$, $\lambda _{1c}\gg 1$ we use Eqs. (\ref{e38}, \ref
{37}) taken at $X=0$. The solution for the functions $\Psi _{i}$ $\left(
i=0,1,2\right) $ can be sought in the form 
\begin{equation}
\Psi _{i}\left( \lambda _{1d},\lambda _{1c}\right) =\lambda _{1d}R_{i}\left(
z\right) \text{, \ }z=\lambda _{1d}\lambda _{1c}  \label{e60}
\end{equation}
Then, we obtain the following equation for $R_{i}$%
\begin{equation}
\left( {\cal H}_{o}+\tilde{\omega}z\right) R_{0}\left( z\right) =0
\label{e60a}
\end{equation}

\begin{equation}
\left( {\cal H}_{o}+\tilde{\omega}z\right) R_{1}\left( z\right) +2\tilde{%
\omega}zR_{0}\left( z\right) =0  \label{e61}
\end{equation}

\begin{equation}
\left( {\cal H}_{o}+\tilde{\omega}z\right) R_{2}-4\tilde{\omega}zR_{1}\left(
z\right) =0  \label{e61a}
\end{equation}
where 
\begin{equation}
{\cal H}_{o}=-2z^{2}\frac{d^{2}}{dz^{2}}  \label{e62}
\end{equation}
Eq. (\ref{e60a}) is exactly the final equation obtained for the orthogonal
ensemble in the standard parametrization. Writing the eigenfunction of the
ground state $\Psi $ in the form 
\begin{equation}
\Psi =\lambda _{1d}R  \label{e62a}
\end{equation}
we find for $R$%
\begin{equation}
R=R_{0}\left( z\left( 1+2\left( \eta \eta ^{\ast }-\kappa \kappa ^{\ast
}\right) \right) \right)  \label{e63}
\end{equation}

In the opposite limit $X^{2}/\tilde{\omega}\rightarrow \infty ,$ the
solution of Eqs. (\ref{e38}, \ref{37}) is completely different. Putting $%
\tilde{\omega}=0$ in this equations we find a solution 
\begin{equation}
\Psi _{0}\left( \lambda _{1c}\right) =\left( 1+X\lambda _{1c}\right) \exp
\left( -X\lambda _{1c}\right)  \label{e64}
\end{equation}
\[
\Psi _{1}=\Psi _{2}=0 
\]
Formally, one could write non-zero solutions for $\Psi _{1}$ and $\Psi _{2}$
proportional to $\Psi _{0}$. However, we understand that in the magnetic
parametrization the ground state function $\Psi $ should not contain at $%
\omega =0$ Grassmann variables because the effective Hamiltonian does not
contain them.

So, in the limit $\omega \rightarrow 0$ the form of the ground state
function $\Psi $ changes discontinuously from Eqs. (\ref{e62a}) to Eq. (\ref
{e64}) as a magnetic field is applied. Physically, it may correspond to
lifting degeneracies of some states of the initial electron problem by the
magnetic field.

The ground state solution, Eq. (\ref{e64}), is written for $\lambda _{1d}$, $%
\lambda _{1c}\gg 1$. We have checked that this state survives and does not
considerably change its form solving Eqs. (\ref{38b}) in the limit $\omega
=0 $. The limit of the vanishing frequency considerably simplifies the
equations because the ``diffuson variable'' $\lambda _{1d}$ decouples from
the ``cooperon variables'' $\lambda _{1c}$ and $\lambda _{c}$. Then, one can
search for a solution $\Psi $ depending on the cooperon variables only. We
used the standard over-relaxation method with the Chebyshev acceleration to
found numerically that up to $X$ as large as $0.3$, Eq.~(\ref{e64}),
describes the solution rather well. At higher values of $X$, no dramatic
changes occurred to this state either.

It is clear that in the limit $X\rightarrow \infty $ essential values of the
cooperon variables $\lambda _{1c}$ and $\lambda _{c}$ are close to unity for
an arbitrary $\omega $ and one obtains the function $\Psi $ for the unitary
ensemble. The solution found at $\omega =0$ can be used for non-zero
frequencies provided the variable $\lambda _{1d}$ is not very large. From
Eqs. (\ref{e38},\ref{37}) we can estimate the region of the applicability of
the solution as 
\begin{equation}
\lambda _{1d}\lambda _{1c}\tilde{\omega}\ll 1  \label{e65}
\end{equation}

In order to find the density-density correlation function $p_{\infty }\left(
x-x^{\prime }\right) $ and its moments $p_{\infty }^{\left( n\right) }\left(
x-x^{\prime }\right) $ at large distances we have to find not only the
ground state but also at least the first excited state $\phi _{1}$. After
that one should calculate the proper ``matrix elements'' in Eq. (\ref{29})
or use Eq. (\ref{27a}). Although we argue that the eigenvalue ${\cal E}_{%
{\rm u}}=1/4$ of the lowest excited state is exact, we are not able, as
previously, to write the proper wave function in all regions of the
variables exactly. The solution for $\phi _{1}$ can be sought in the form,
Eq. (\ref{34}), corresponding to the symmetry of the block $Q^{12}$ of the
supermatrix $Q$, which is the usual way of constructing the proper excited
eigenstates \cite{efetov,book}. Explicit formulae can be written at $X\ll 1$
provided the variables $\lambda _{1d}$, $\lambda _{1c}$ are large, $\lambda
_{1d}$, $\lambda _{1d}\gg 1$, but limited from above by the inequality (\ref
{e65}). In this limit one may keep the function $f_{1}$ only and we come to
the following equation 
\begin{equation}
\left( {\cal H}_{X}^{\left( 0\right) }-2\lambda _{1d}\frac{\partial }{%
\partial \lambda _{1d}}\right) f_{1\rho }={\cal E}\left( \rho \right)
f_{1\rho }  \label{e66}
\end{equation}
${\cal H}_{X}^{\left( 0\right) }$ is given by Eq. (\ref{37}) and $\rho $ is
a continuous real variable. The variables $\lambda _{1d}$ and $\lambda _{1c}$
in Eq. (\ref{e66}) separate from each other and we write the solution $f_{1}$
(see Eq. (\ref{34}) in the form 
\begin{equation}
f_{1\rho }=\lambda _{1d}^{-1/2}\chi _{\rho }\left( \ln \lambda _{1d}\right)
\Psi _{0}\left( \lambda _{1c}\right)   \label{e67}
\end{equation}
with the function $\Psi _{0}$, given by Eq. (\ref{e64}). The function $\chi
\left( v\right) $ should be written as 
\begin{equation}
\chi _{\rho }\left( v\right) =\left( \frac{2}{\pi }\right) ^{1/2}\sin \left[
-\left( \rho /2\right) v+\delta \left( \rho \right) \right]   \label{e67b}
\end{equation}
Eq. (\ref{e67b}) corresponds to a picture of ``free motion'' suggested in
Ref. \cite{book} for a description of the region of not very large $\lambda
_{1}$. In the case considered here, Eq. (\ref{e67}), is applicable in the
region determined by the inequality (\ref{e65}). In calculation of
asymptotics of the density-density correlator at large distances small
values of $\rho \ll 1$ give the main contribution and, in this limit, the
phase $\delta $ is also small, $\delta \left( \rho \right) \sim \rho $ (see
Ref. \cite{book}).

From Eq. (\ref{e67}) we obtain immediately 
\begin{equation}
{\cal E}\left( \rho \right) =\frac{1}{4}\left( 1+\rho ^{2}\right)
\label{e67a}
\end{equation}

The form of the solution $f_{1}$, Eq. (\ref{e67}), differs from the ground
state by a dependence on the diffuson variables $\lambda _{1d}$. This is the
same difference as the one for the unitary ensemble. It is not difficult to
check that the variable $\lambda _{1d\text{ }}$ separates from the cooperon
variables even if $\lambda _{1c}$ is not large, $\lambda _{1c}\sim 1$. Then,
the solution (\ref{e67}) may still be used in the region specified Eq. (\ref
{e65}) provided the function $\Psi _{0}\left( \lambda _{1c}\right) $ is
replaced by the solution $\Psi $ (depending only on the cooperon variables)
of Eqs. (\ref{38a}, \ref{38b}) taken at $\omega =0$.

Eq. (\ref{e67}) with the function $\Psi $ can be used also for an arbitrary $%
X$. In the limit $X\gg 1$, the function $\Psi $ decays fast as the variables 
$\lambda _{1c}$ and $\lambda _{c}$ deviate from $1$. This means that the
cooperon variables are `frozen' and we come to the unitary ensemble. Then,
the restriction (\ref{e65}) is not important and all formulae can be written
for arbitrary $\lambda _{1d}\omega $ \cite{efetov,book}. Of course, in order
to prove the existence of the states with the eigenvalues ${\cal E}\left(
\rho \right) $, (\ref{e67}), rigorously one should investigate the exact
equations for arbitrary values of the cooperon and diffuson variables and
prove that the solutions `behave well' everywhere. It is a difficult task
that can apparently be performed only numerically.

Assuming nevertheless that this state exists and is given for $\lambda _{1d}$%
, $\lambda _{1c}\gg 1$ by Eq. (\ref{e67}) (the inequality (\ref{e65}) should
also be fulfilled) we can estimate the proper matrix elements, Eq. (\ref{29}%
), determining the average density-density correlation function $p_{\infty
}\left( x\right) $ and its moments $p_{\infty }^{\left( n\right) }\left(
x\right) $, Eq. (\ref{00}). However, trying to calculate the integrals in
Eq. (\ref{29}) one encounters a difficulty. The integrals over the matrix
elements of $Q$ can be reduced to integrals over the variables $\lambda
_{1d} $, $\lambda _{1c},\lambda _{d}$. $\lambda _{c}$ and the Grassmann
variables. Proceeding in this way one should use the Berezinian $J$ of the
transformation to these variables, Eq. (\ref{35}), which is singular at $%
\lambda _{1c}=\lambda _{c}=1$. The singularity in the Berezinian leads
usually to additional contributions. To avoid explicit calculations of these
contributions it is more convenient to calculate not the physical quantities
of interest themselves but their derivative over $X$. This leads to
additional factors $\lambda _{1c}^{2}-\lambda _{c}^{2}$ in integrands, thus
compensating the singularities of the Berezinians.

Calculation of the physical quantities using the spectral expansion, Eq. (%
\ref{29}), is still very difficult. The alternative way of calculations is
solving Eq. (\ref{26}, and calculating the integral in Eq. (\ref{27a}). A
simplified differential form of Eq. (\ref{26}) can be written for the
function $P_{k}^{\left( n\right) }$ determining the correlator $p_{\infty
}^{\left( n\right) }$ as 
\begin{equation}
{\cal H}_{X}^{\left( 0\right) }P_{k}^{(n)}+i\kappa P_{k}^{(n)}=\lambda
_{1d}^{n}\lambda _{1c}^{n}\,\Psi _{0}\left( \lambda _{1c}\right)  \label{e68}
\end{equation}

where $\kappa =kL_{{\rm cu}}$. In Eq. (\ref{e68}), we do not write Grassmann
variables explicitly implying that $P^{\left( n\right) }$ is the non-compact
central part similar to $\left( 1-\lambda _{1c}^{2}\right) ^{1/2}f_{1}$ in
Eq. (\ref{34}).

The derivative over the field $\left( T_{nn}\right) _{X}^{\prime }$ of the
correlator $T_{11},$ Eq. (\ref{27a}) can be reduced to the form 
\begin{equation}
\frac{dT_{nn}\left( k\right) }{dX}\sim \frac{d}{dX}\int \lambda
_{1d}^{n-2}\lambda _{1c}^{n-4}\tilde{P}_{k33}^{\left( n\right) }d\lambda
_{1d}d\lambda _{1c}  \label{e69}
\end{equation}
Writing Eq. (\ref{e69}) the Berezinian $J$, Eq. (\ref{35}), was used in the
limit $\lambda _{1c}$, $\lambda _{1d}\gg 1$ and $\tilde{P}_{k}^{\left(
n\right) }=P_{k}^{\left( n\right) }+P_{-k}^{\left( n\right) }$.

Using the fact that the Hamiltonian ${\cal H}_{X}^{\left( 0\right) }$
contains the variables $\lambda _{1d}$ and $\lambda _{1c}$ separately let us
expand the function $P_{k}^{\left( n\right) }\left( \lambda _{1d},\lambda
_{1c}\right) $ in the eigenfunctions of the diffuson part of the
Hamiltonian. Writing 
\begin{equation}
P_{k}^{\left( n\right) }\left( \lambda _{1d},\lambda _{1c}\right)
=\sum_{\rho }\lambda _{1d}^{1/2}\chi _{\rho }\left( \ln \lambda _{1d}\right)
P_{k\rho }^{\left( n\right) }\left( \lambda _{1c}\right)  \label{e70}
\end{equation}
we obtain the following equation for $P_{k\rho }^{\left( n\right) }\left(
\lambda _{1c}\right) $%
\begin{equation}
\left( {\cal E}\left( \rho \right) +i\kappa \right) P_{k\rho }^{\left(
n\right) }\left( \lambda _{1c}\right) +\tilde{H}_{X}^{\left( 0\right)
}P_{k\rho }^{\left( n\right) }\left( \lambda _{1c}\right) =A\left( \lambda
_{1c}\right)  \label{e71}
\end{equation}
where 
\[
A\left( \lambda _{1c}\right) =\int \lambda _{1d}^{n-3/2}\chi _{\rho }\left(
\ln \lambda _{1d}\right) \lambda _{1c}^{n}\Psi _{0}\left( \lambda
_{1c}\right) d\lambda _{1d} 
\]
The operator $\tilde{H}_{X}^{\left( 0\right) }$ is the part of the
Hamiltonian ${\cal H}_{X}^{\left( 0\right) }$ in Eq. (\ref{37}) acting on
the variables $\lambda _{1c}$.

A similar expansion was done in Ref. \cite{book} to consider the case of a
strong disorder. The main contribution in the integral over $\lambda _{1d}$
comes from large $\lambda _{1d}$ and the integral formally diverges. When
considering the unitary ensemble one had to cutoff the integral by $\lambda
_{1c}\sim 1/\tilde{\omega}$. At weak magnetic fields considered now, the
cutoff can be determined from Eqs. (\ref{21}, \ref{33}). We see that the
term with the frequency can be neglected if $\lambda _{1d}\lambda _{1c}%
\tilde{\omega}\ll 1$, which leads to the cutoff $\lambda _{1d}\sim \left(
\lambda _{1c}\tilde{\omega}\right) ^{-1}$. So, we estimate the function $%
A\left( \lambda _{1c}\right) $ at $\rho \ll 1$ as 
\begin{equation}
A\left( \lambda _{1c}\right) \sim \left( -i\tilde{\omega}\right)
^{1/2-n}\rho \lambda _{1c}^{1/2}\Psi _{0}\left( \lambda _{1c}\right)
\label{e72}
\end{equation}

\bigskip In a similar way we can reduce Eq. (\ref{e69}) to the form 
\begin{equation}
\frac{dT_{nn}\left( k\right) }{dX}\sim \left( -i\tilde{\omega}\right)
^{1/2-n}\frac{d}{dX}\int P_{k\rho }^{\left( n\right) }\left( \lambda
_{1c}\right) \Psi _{0}\left( \lambda _{1c}\right) \frac{d\lambda _{1c}}{%
\lambda _{1c}^{7/2}}  \label{e73}
\end{equation}
Now, we have to solve Eq. (\ref{e71}) and calculate the integral, Eq. (\ref
{e73}). In order to solve Eq. (\ref{e71}) we introduce a Green function $g$
of this equation 
\begin{equation}
\left( {\cal E}\left( \rho \right) +i\kappa \right) g_{k\rho }\left( \lambda
_{1c},\lambda _{1c}^{\prime }\right) +\tilde{H}_{X}^{\left( 0\right)
}g_{k\rho }\left( \lambda _{1c},\lambda _{1c}^{\prime }\right) =\lambda
_{1c}^{4}\delta \left( \lambda _{1c}-\lambda _{1c}^{\prime }\right)
\label{e74}
\end{equation}
and write the solution as 
\begin{equation}
P_{k\rho }^{\left( n\right) }=\int g_{k\rho }\left( \lambda _{1c},\lambda
_{1c}^{\prime }\right) A\left( \lambda _{1c}^{\prime }\right) \frac{d\lambda
_{1c}^{\prime }}{\left( \lambda _{1c}^{\prime }\right) ^{4}}  \label{e75}
\end{equation}
Then, Eq. (\ref{e73}) is reduced to the form 
\begin{equation}
\frac{dT_{nn}\left( k\right) }{dX}\sim \left( -i\tilde{\omega}\right) ^{1-2n}%
\frac{d}{dX}\int g_{k\rho }\left( \lambda _{1c},\lambda _{1c}^{\prime
}\right) \Psi _{0}\left( \lambda _{1c}\right) \Psi _{0}\left( \lambda
_{1c}^{\prime }\right) \frac{\rho ^{2}d\rho d\lambda _{1c}d\lambda
_{1c}^{\prime }}{\left( \lambda _{1c}\lambda _{1c}^{\prime }\right) ^{7/2}}
\label{e76}
\end{equation}
The last step to be done is to find the function $g$ and calculate the
integral in Eq. (\ref{e76}). As usual, \cite{Gogol,book} the combination $%
{\cal E}\left( \rho \right) +i\kappa $ enters the final formulae and one
should perform Fourier transform to get the coordinate dependence. The point 
$\kappa =i{\cal E}\left( \rho \right) $ is a branching point and the
integral over $\kappa $ can be shifted such that the integration is
performed along the edges of the cut $\left( i{\cal E}\left( \rho \right)
,i\infty \right) $. The point $i{\cal E}\left( 0\right) =i/4$ gives the
value of the localization length whereas the integration over $\rho $ leads
to an additional power law prefactor $\left| x\right| ^{-3/2}$. Being
interested mainly in determining the exponential decay we can immediately
understand that all correlators $T_{nn}$ decay as $\exp \left( -\left|
x\right| /4L_{{\rm cu}}\right) $. To calculate the prefactor we may put $%
\kappa =i{\cal E}\left( \rho \right) $ in Eq. (\ref{e74}) and find the
proper solution $g_{0}\left( \lambda _{1c},\lambda _{1c}^{\prime }\right) $,
which satisfies the equation 
\begin{equation}
\tilde{H}_{X}^{\left( 0\right) }g_{0}\left( \lambda _{1c},\lambda
_{1c}^{\prime }\right) =\lambda _{1c}^{4}\delta \left( \lambda _{1c}-\lambda
_{1c}^{\prime }\right)  \label{e77}
\end{equation}
and has to be substituted into Eq. (\ref{e76}).

The solution of Eq. (\ref{e77}) can be found easily because one of the
solutions $\Psi _{0}\left( \lambda _{1c}\right) $, Eq. (\ref{e64}), of the
corresponding homogeneous equation is known. We can write immediately the
second solution $\Phi _{0}\left( \lambda _{1c}\right) $ in the form 
\begin{equation}
\Phi _{0}\left( \lambda _{1c}\right) =\Psi _{0}\left( \lambda _{1c}\right)
\int_{0}^{\lambda _{1c}}\frac{\lambda ^{2}}{\Psi _{0}^{2}\left( \lambda
\right) }d\lambda  \label{e78}
\end{equation}
The function $\Phi _{0}\left( \lambda _{1c}\right) $ grows exponentially at $%
\lambda _{1c}\rightarrow \infty $ and is proportional to $\lambda _{1c}^{3}$
for small $\lambda _{1c}$. Using the functions $\Psi _{0}\left( \lambda
_{1c}\right) $ and $\Phi _{0}\left( \lambda _{1c}\right) $ we write the
Green function $g_{0}\left( \lambda _{1c},\lambda _{1c}^{\prime }\right) $
as 
\begin{equation}
g_{0}\left( \lambda _{1c},\lambda _{1c}^{\prime }\right) =\left\{ 
\begin{array}{cc}
\Phi _{0}\left( \lambda _{1c}\right) \Psi _{0}\left( \lambda _{1c}^{\prime
}\right) , & \lambda _{1c}\leqslant \lambda _{1c}^{\prime } \\ 
\Psi _{0}\left( \lambda _{1c}\right) \Phi _{0}\left( \lambda _{1c}^{\prime
}\right) , & \lambda _{1c}\geqslant \lambda _{1c}^{\prime }
\end{array}
\right.  \label{e79}
\end{equation}

\ Substituting the function $g_{0}$, Eqs. (\ref{e79}, \ref{e64}, \ref{e78}),
into Eq. (\ref{e76}) we see that in the region $1\ll \lambda _{1c},\lambda
_{1c}^{\prime }\ll 1/X$ the integral is logarithmic and is safely cut from
below and above by the limits of this region. It is important that we remain
in the region of large $\lambda _{1c}$ where our approximations are valid.
As a result of the calculation we obtain for the moments $p_{\infty
}^{\left( n\right) }\left( x\right) $ of the density-density correlation
function 
\begin{equation}
\frac{dp_{\infty }^{\left( n\right) }\left( x\right) }{dX}\sim X\ln \left(
1/X\right) \left| x\right| ^{-3/2}\exp \left( -\left| x\right| /4L_{{\rm cu}%
}\right)  \label{e80}
\end{equation}
It is interesting to note that the same prefactor determines correlations at
coinciding points. Writing $p_{\infty }^{\left( n\right) }\left( 0\right) $
as 
\[
\frac{dp_{\infty }^{\left( n\right) }\left( 0\right) }{dX}\sim \left( -i%
\tilde{\omega}\right) ^{2n-1}\int \frac{\left( \lambda _{1d}\lambda
_{1c}\right) ^{2n}d\lambda _{1d}d\lambda _{1c}}{\lambda _{1d}^{2}\lambda
_{1c}^{2}} 
\]
and integrating as previously over the region $0<\lambda _{1d}\leq \left(
\lambda _{1c}\tilde{\omega}\right) ^{-1}$, $1\leq \lambda _{1c}\leq 1/X$ we
come to the result 
\begin{equation}
\frac{dp_{\infty }^{\left( n\right) }\left( 0\right) }{dX}\sim X\ln \left(
1/X\right)  \label{e81}
\end{equation}

Eq. (\ref{e80}) shows that the derivative over the magnetic field of all
moments of the density-density correlation function decays with the
localization length $L_{{\rm cu}}$ of the unitary ensemble. The moments
themselves can be obtained by integrating Eq. (\ref{e80}) over the magnetic
field. The result can be written as 
\begin{equation}
p_{\infty }^{(n)}(x)\propto \left( \frac{1}{x}\right) ^{3/2}\left[ \exp
\left( -\frac{\left| x\right| }{4L_{{\rm co}}}\right) +C_{n}X^{2}\ln \left( 
\frac{1}{X}\right) \exp \left( -\frac{\left| x\right| }{4L_{{\rm cu}}}%
\right) \right] \;.  \label{62}
\end{equation}
where $C_{n}$ is a coefficient depending on $n$. The first term in Eq. (\ref
{62}) is added assuming that at zero magnetic field one should have the
standard result for the orthogonal ensemble. (In our previous work \cite{WE}
we wrote $\ln ^{2}\left( 1/X\right) $ instead of $\ln \left( 1/X\right) $
which was a result of a not sufficiently accurate estimate). The second term
in Eq. (\ref{62}) becomes larger than the first one at distances exceeding
\begin{equation}
x_{H}\sim L_{c}\ln \left( 1/X\right)  \label{e100}
\end{equation}
and decays in the same way for all the moments. As we have discussed in the
previous Section, this should mean that this term is determined by rare
splashes of the wave functions. These splashes can be due to a hybridization
of states with eigenenergies very close to each other and it is not
surprising that they are sensitive to the magnetic field.

A more delicate question concerns the distribution of the logarithm of the
wave functions, governing the behavior of ``typical'' wave functions. This
quantity is necessary for comparison with a recent numerical study, \cite
{Been-Sch} where the logarithm of the transmittance was shown to behave
smoothly between the orthogonal and unitary ensembles. 

Our new state with the eigenenergy ${\cal E=}\left( 1+\rho ^{2}\right) /4$
contributes to the distribution of the logarithms also. Unfortunately, the
procedure developed for calculation of the moments cannot be extended to a
calculation of the distribution function in the most interesting relation of
the variables $t\approx \exp \left( -\left| x\right| /L_{{\rm cu}}\right) $
which would give the distribution of the localization lengths. The problem
is that one needs to know the analytical properties of the Green function 
$g_{k\rho }$, Eq. (\ref{e74}), for arbitrary complex $\rho $. One can 
understand this on the example of the pure ensembles. In Eq. (\ref{59}), 
computation of the distribution function required shifting the contour of 
the integration over $\rho $ to the saddle point $\rho =-i\ln t/u\approx i$. 
One could perform this procedure there because the integrand was known.

In the crossover regime, one can rely only on the picture of the ``free
motion'' applicable at small $\rho $, which make calculation of the
distribution function not feasible. We can say only that the distribution
function has an additional weight at $t\approx \exp \left( -\left| x\right|
/L_{{\rm cu}}\right) $ so that the entire function is broader that for the 
pure ensembles. At the same time, the main body of the function is not known, 
which does not allow us to say much about ``typical'' wave functions. So, 
the results that we really have do not seem to contradict to the results of 
the numerical study of the average logarithms \cite{Been-Sch}, where the 
second length was not seen. They suggest also that the information the 
Borland conjecture refers to, the behavior of the averaged logarithm of the 
wave function, may be not sufficient to characterize the crossover. 
Instead, the entire distribution function is of interest since it can differ 
from the standard logarithmically normal even for $x \gg L_c$.

As concerns a numerical observation of the second length $L_{{\rm cu}}$
in the moments of transmittances or density-density correlations, it may be
very difficult because one needs to make computation for a very large number
of configurations to be able to make a reliable statistics of the rare
events. In addition, a smearing due to the presence of leads can make the
observation even more difficult.

\section{Discussion of the Results}

We see from Eq.~(\ref{62}) that for weak magnetic fields the first term
dominates at small distances. However, whatever weak the magnetic field is,
the second term always prevails at sufficiently large distances. This is due
to the fact that, although the small crossover parameter $X$ enters the
pre-exponential factor, the exponential function containing the localization
length $L_{{\rm cu}}=2L_{{\rm co}}$ decays much slower. Comparing these two
terms, we obtained the characteristic distance $x_{H}$, Eq. (\ref{e100})
where the both terms are of the same order of magnitude

To explain Eq. (\ref{62}) qualitatively one can follow argument suggested by
Mott for describing the AC conductivity \cite{Mott2,Mott3} with minor
modifications necessary in our case.

Actually, the arguments presented below apply for localized states in any
dimensionality and allow us to obtain the scale $x_{H}$, Eq.~(\ref{e100}),
without any calculation. Naturally, the localization length $L_{c}$ in Eq.~(%
\ref{e100}) depends on the dimensionality of the space. However, to be
precise, we will speak in terms of the localization in wires.

In quasi-one-dimensional samples all states are localized (in other words, $%
\int_{-\infty }^{+\infty }p_{\infty }^{(1)}(x)dx=1$). At the same time, for
frequencies exceeding the energy $E_{c}=D_{0}/L_{c}^{2}$ the system behaves
like a metal. Therefore, eigenenergies of states localized in a region of
the wire of size $L_{c}$ are separated by the mean level spacing of the
order of $E_{c}$. This picture allows us to estimate immediately the
magnetic field $H_{c}$ at which the crossover between the orthogonal and
unitary ensembles occurs. One should simply consider the region of the size $%
L_{c}$ as a closed grain and recall that the crossover between the
orthogonal and unitary ensembles in a metallic grain occurs \cite
{efetov,book} at flux $\phi $ through the grain given by the relation 
\begin{equation}
E_{T}\left( \phi /\phi _{0}\right) ^{2}\sim \Delta _{G}\;,  \label{a2000}
\end{equation}
where $\phi _{0}$ is the flux quantum, $\Delta _{G}=\left( \nu V\right)
^{-1} $ is the mean level spacing in the grain ($V$ is the volume), having
the same order of magnitude as the Thouless energy $E_{T}=D_{0}/L^{2}$. For
a quasi-one-dimensional grain of length $L=L_{c}$ one obtains the
characteristic value $X\sim 1$, which means that the magnetic flux through
the segment of the length $L_{c}$ is equal to the flux quantum $\phi _{0}$ 
\cite{EL}. At such fields the localization length of the main body of the
wave functions changes considerably and this was clearly seen in the
numerical work. \cite{Been-Sch}

At the same time, we know \cite{EfAl,book} that the low frequency
asymptotics of the level-level correlation function in an isolated grain
changes from $\omega $ to $\omega ^{2}$ as soon as an arbitrarily weak
magnetic field is applied. This corresponds to a finite probability of
having two different levels at any small distance $\omega$. These levels
hybridize by the magnetic field and this leads to the low frequency
asymptotics $\omega ^{2}$ typical for the unitary ensemble.

The two-scale localization considered in the previous sections is of a
similar origin, namely, one can always find two localized states that are
arbitrarily close to each other in energy, although they may be separated by
a large distance in space. Using this fact Mott \cite{Mott2,Mott3}
calculated the AC conductivity at finite frequencies $\omega $. We do not
study the conductivity itself but rather different kinetic quantities like
e.g. the density-density correlation function. In our case, the role of an
external perturbation is taken over by the magnetic field.

Following Mott's arguments, \cite{Mott2,Mott3} let us try to find states
whose energies slightly differ from the energy of a chosen state. As we have
mentioned above, the typical separation in energy of states localized within
a region of the length $L_{c}$ is of order $E_{c}$. In neighboring regions
of the lengths $L_{c}$ one cannot find such states either: Even if two
states were occasionally closely located in two initially isolated systems,
they would hybridize and split after bringing these two segments of the wire
in contact. The value of the splitting would be $E_{c}$ again. However, if
two states are separated by a large distance $x\gg L_{c}$, the splitting
energy $\Delta _{x}$ decays exponentially 
\begin{equation}
\Delta _{x}\sim E_{c}\exp (-x/L_{c})  \label{a2001}
\end{equation}
due to exponentially decaying overlap between the localized states. (Mott
considered a tight-binding model; hence, the energy characterizing the
splitting in his consideration was $\Delta _{x}\sim I_{0}\exp (-x/L_{c})$,
where $I_{0}$ is of the order of band width.)

Thus, one can find states $a$ and $b$ with an exponentially small energy
difference considering states localized sufficiently far from each other.
Now, turning on an external perturbation (magnetic field in our case) we
want to mix these states, which, like in isolated granules, would lead to a
behavior typical for the unitary ensemble (with the doubled localization
length $L_{{\rm cu}}=2L_{{\rm co}}$). However, in order to mix the states we
need a sufficiently large matrix element $A_{ab}$ of the vector potential
between the states. If the states did not hybridize at all this matrix
element would be exponentially small and this is similar the situation one
encounters when calculating the AC conductivity at low frequencies. Mott 
\cite{Mott2,Mott3} suggested to take into account the states when they start
hybridizing and we use this idea for our estimates. In this case the matrix
element $A_{ab}$ is not exponentially small because the hybridized wave
function is concentrated equally in the both localization centers.

Using the modelling of the localization center in terms of an isolated
granule of the length $L_{c}$ and applying the magnetic field $H$ we obtain
easily a characteristic energy window $E_{H}$ of the order of 
\begin{equation}
E_{H}\sim X^{2}E_{c}\;.  \label{EH}
\end{equation}
This estimate follows from Eq.~(\ref{21}) by using the fact that in wires
the energies $E_{c}$, $E_{T}$ and $\Delta _{G}$ are of the same order. In an
isolated grain states within this window are mixed and their correlations
obey the statistics of the unitary ensemble.

So, we come to the following picture:

If two localized states $a$ and $b$ are separated by a distance $x$ such
that the corresponding overlap integral $\Delta _{x}$ exceeds $E_{H}$, the
states are not considerably affected by the magnetic field and its influence
can be considered perturbatively.

If two localized states $a$ and $b$ are separated by a very large distance $%
x $ such that the overlap $\Delta _{x}$ is much smaller than $E_{H}$, then
the matrix $A_{ab}$ is exponentially small and the effect of the magnetic
field on the states can be neglected.

The magnetic field is important if the distance $x$ between the localized
states $a$ and $b$ is such that $\Delta _{x}\sim E_{H}$. For such states the
matrix element $A_{ab}$ is not small but, at the same time, the levels are
close to each other. The magnetic field $H$ mixes the states $a$ and $b$ and
one can expect for the hybridized states the statistics of the unitary
ensemble. Comparing the energies $\Delta _{x}$ and $E_{H}$%
\begin{equation}
\exp (-x/L_{c})\sim X^{2}\;,  \label{72}
\end{equation}
we obtain the characteristic distance $x_{H}$, Eq.~(\ref{e100}).

Assuming that the mixing of the levels by the magnetic field leads to the
correlations given by the unitary ensemble we come to the conclusion that
the exponential decay of the density-density correlation function and other
correlation functions characterizing wave functions should be determined at
distances $x\gtrsim x_{H}$ by the localization length $L_{{\rm cu}}$ of the
unitary ensemble.

These simple arguments explain the two-scale behavior, Eq.~(\ref{62}), and
give the correct second scale $x_{H}$, although the doubling of the
localization length can be obtained by the exact calculation only. At the
same time, the probability to obtain states at a desired distance with
eigenenergies very close to each other is small. This small probability may
be compensated by a large contribution to moments of the density and
therefore can be essential when calculating these quantities. As concerns
calculation of the logarithm the large splashes are not crucial and their
contribution is small due to the small probability, which is in agreement
with the results of the numerical investigation of the transmittance. \cite
{Been-Sch}(actually, the authors of this work suggested as one of possible
explanation of the difference between the averaged logarithm and the moments
existence of ``anomalously'' localized states).

An additional suppression of the splashes can be a consequence of a smearing
due to inelastic scattering and presence of leads. To observe the splashes
one has to be at distances from the ends of the wire exceeding $x_{H}$. At
distances smaller that $x_{H}$ the weak magnetic field cannot have any
considerable effect of the wave functions because the validity of the
perturbation theory in $H$ is determined by the value of the smearing and
not by the energy separation of the relevant states. But the transmittance
probes the wave functions at the end of the sample. In this region, the
magnetic field can become important only if the relevant magnetic energy $%
E_{H}$ becomes of the order of the smearing energy and the latter is of the
order of $E_{c}$ near the ends.

Recently, localization in disordered wires in a magnetic field was studied
analyzing the autocorrelation function of spectral determinants (ASD) \cite
{Kettemann} and no indication of the two-scale localization has been found.
In our opinion, this cannot be a great surprise because the ASD is a
quantity that can give only a rough picture of what is going in reality. By
definition, the ASD probes only correlations of the energy levels and not
the spatial structure of the wave functions. Any localized states contribute
to the ASD independent of the distance between the centers of localization.
This makes hardly possible to say anything about the overlap of the
localized states, which is so important for the existence of the second
scale. We think that the fact that the second scale was not identified in
Ref. \cite{Kettemann} is not in contradiction with our finding but rather
demonstrates that the ASD is not a proper quantity to describe the effect
involved.

Let us now discuss how one can try to check experimentally our predictions.
Considering the transmittance of single wire one should speak about the
logarithm of the transmittance that is a self-averaging quantity. In this
case, the rare splashes we discussed previously are not important and one
should see only one localization length which is at weak magnetic fields
close to the length $L_{{\rm co}}$ of the orthogonal ensemble. However, one
can measure the conductance of a system of a large number of wires connected
in parallel. This is exactly the set up of the experiment of Ref. \cite
{Exp1,Exp2}. The large number of wires in this case leads to averaging of
the transmittance itself rather than of its logarithm. In such a situation
one might see, in principle, the second scale $L_{{\rm cu}}$. Physically
this means that although the splashes are rare, a wire, where such a splash
occurs, gives the main contribution to the conductivity.

Unfortunately, in the experiment \cite{Exp1,Exp2}, the localization length
was not measured directly but was extracted from the dependence of the
hopping conductivity on temperature.

It is known that the conductance at very low temperatures is possible due to
activations through Mott resonances, which leads to the picture of the
variable range hopping. \cite{Mott2} The localization length $L_{c}$ enters
directly the activation energy in this conductivity 
\begin{equation}
\sigma =\sigma _{0}\exp \left[ \left( -T_{0}/T\right) ^{1/(1+d)}\right] 
\text{, \ \ \ }T_{0}\sim \left( \nu L_{c}^{d}\right) ^{-1}\;.  \label{hopp1}
\end{equation}
where $d$ is the dimensionality of the system. However, the localization
length entering this formula is the length for a typical wave function and
the splashes are not very important.

As a matter of fact, Eq.~(\ref{hopp1}) does not apply to 1D samples. The
conductivity of 1D chains and wires should obey a simple Arrhenius-type law 
\cite{kurk} instead 
\begin{equation}
\sigma =\sigma _{0}\exp \left( -T_{0}/T\right) \text{, \ \ \ }T_{0}\sim
\left( \nu SL_{c}\right) ^{-1}\;.  \label{hopp2}
\end{equation}
Remarkably, even accounting for electron-electron interactions preserves
this kind of dependence. \cite{larkin} It is this law that was discovered in
recent experiments in the regime of strong localization \cite{Exp1,Exp2} and
activation energy $T_{0}$ was used to extract the localization length $L_{c}$%
. Which localization length enters the conductivity in the $1D$ case is not
as clear as for higher dimensions and there could be a chance that the
second scale can be seen here. Performing such experimental analysis at
different temperatures one could try to observe the doubling of the
localization length considered, because decreasing the temperature should
result in a crossover from the activation behavior with the activation
energy $T_{0}$ in Eq. ~(\ref{hopp2}), to the activation energy $T_{0}/2$.
Indeed, at very low temperatures, $T<\left( \nu Sx_{H}\right) ^{-1}$, the
electron hopping due to the overlap of far tails plays the dominant role, so
that the larger localization length $L_{{\rm cu}}$ should be used in Eq. (%
\ref{hopp2}). For temperatures $\left( \nu Sx_{H}\right) ^{-1}<T<\left( \nu
SL_{c}\right) ^{-1}$, hopping between the main body of localized states is
essential and the localization length $L_{{\rm co}}$ should be substituted
for $L_{c}$ in Eq.~(\ref{hopp2}).

Needless to say that the second scale may be observed only if the wires are
sufficiently long such that the main contribution to the resistivity comes
from the variable-range hopping inside the sample and the wire can be
considered effectively as infinite. In this limit the resistivity should be
proportional to the length of the wire. Measuring the conductance of a
finite wire at ultralow temperatures where the resistivity grows
exponentially with the sample length cannot help extracting the two-scale
behavior as it has been discussed previously.

\section{Conclusions}

Studying localization properties of infinitely long disordered wires within
supersymmetry technique combined with the transfer-matrix technique, we
managed to calculate correlation functions of interest. This procedure is
based on working in the coordinate representation, Eq.~(\ref{16}), and
generating any quantity of interest from the two-point correlation function,
Eq.~(\ref{13}). Using this approach we obtained all moments and distribution
functions of the density-density correlator and Landauer-like conductivity
for the orthogonal and unitary ensembles, Eqs.~(\ref{54},\ref{55},\ref{58}).
These expressions are exact and they are valid for arbitrary distances.

Studying the crossover between the unitary and orthogonal ensembles for weak
magnetic fields we have found moments of the density-density correlations at
large distances $x\gg L_{c}$. We have demonstrated that the far tail of
these correlation functions characterizing the wave functions decays at
arbitrary weak magnetic field with the localization length which is double
as large as that of the main part. As all the moments decay with the same
length, we argue that this behavior is due to rare splashes of the wave
functions that are very sensitive to the magnetic field. In contrast to
this, the localization length characterizing a typical wave function might
exhibit a smooth crossover.

We emphasize, however, that in the crossover regime, not only study of the
typical localization length, but also of the whole distribution of the
logarithm is of interest. It should differ from the standard log-normal one
even for $x\gg L_{{\rm c}}$ so long as $x\lesssim x_{H}$, Eq. (\ref{e100}),
being asymmetric and broad. At present, we can not predict how the entire 
distribution function of the averaged logarithm of the wave function will 
develop in the limit of infinitely long distances. 
%As concerns the distance $x$ $\gg x_{H}$ one may apparently expect a 
%log-normal distribution near the localization length $L_{%
%{\rm co}}$ of the orthogonal ensemble. As the magnetic field grows, the peak
%of the distribution may smoothly move toward the localization length $L_{%
%{\rm cu}}=2L_{{\rm co}}$. Unfortunately, at present we cannot describe the
%main body of the distribution function in the crossover regime.

Slightly revising the Mott arguments for the AC conductance, \cite
{Mott2,Mott3} we have shown how one can extract the characteristic scale of
the problem $x_{H}$, without performing any calculations. These arguments
are quite general, which implies that our results may be valid for a
disordered system of any dimension.

From the exact treatment as well as from the qualitative arguments of the
previous section we understand that the effect we have found is very
sensitive to different kinds of the level smearing. Therefore, it may be
observed at low temperatures only.

As concerns a possibility of an experimental observation of the effect it
might be seen at low temperatures in the regime of the hopping conductivity
in a system of wires connected in parallel. The wires should be sufficiently
long, such that the system would obey Ohm's law. Then, if the tails of the
wave functions determine the hopping conductivity, one should expect the
Arrhenius law but with different activation energies in the limit of very
low temperatures.

We hope that it can become possible to check our finding both numerically
and experimentally, although this can be a difficult task due to a small
probability of the wave functions with the high sensitivity to the magnetic
field.

\section{Acknowledgments}

This work was supported by SFB 237 ``{\it Unordnung und Gro{\ss}e
Fluktuationen''}. A.V.K. acknowledges hospitality of the Instituut-Lorentz
at the Leiden University. We are indebted to I.L. Aleiner, B.L. Altshuler,
C.W.J.~Beenakker, M.~Janssen, and H.~Schomerus for useful discussions.

\end{document}